\documentclass{aa}  
\usepackage{graphicx}
\usepackage{natbib}
\usepackage{threeparttable}
\usepackage{txfonts}
\usepackage{appendix}

\def\cmfgen{{\sc cmfgen}}
\def\iraf{{\sc iraf}}
\def\mesa{{\sc mesa}}

\def \lsun{\ifmmode{{\rm\ L}_\odot}\else{${\rm\ L}_\odot $}\fi}

\def \msun{\ifmmode{{\rm\ M}_\odot}\else{${\rm\ M}_\odot$}\fi}

\def \zsun{\ifmmode{{\rm\ Z}_\odot}\else{${\rm\ Z}_\odot$}\fi}

\def \rsun{\ifmmode{{\rm\ R}_\odot}\else{${\rm\ R}_\odot$}\fi}

\def \mdot{\ifmmode{{\rm\dot{M}}}\else{${\rm\dot{M}}$}\fi}

\newcommand\as{${''}$}

\newcommand{\ha}{H$\alpha${}}

\newcommand{\hb}{H$\beta${}}

\newcommand{\hii}{H\,{\sc ii}{}}

\newcommand{\oiii}{[O\,{\sc iii}]}

\newcommand{\nii}{[N\,{\sc ii}]}

\newcommand{\feii}{Fe\,{\sc ii{}}}

\def\lesssim{\mathrel{\hbox{\rlap{\hbox{\lower4pt\hbox{$\sim$}}}\hbox{$<$}}}}
\def\gtrsim{\mathrel{\hbox{\rlap{\hbox{\lower4pt\hbox{$\sim$}}}\hbox{$>$}}}}

\begin{document}

   \title{Type II supernovae as probes of environment metallicity: 
   observations of host \hii\ regions}
	\authorrunning{Anderson et al.}
	\titlerunning{SNe~II as metallicity tracers}
%   \subtitle{SNe~II}

   \author{J. P. Anderson$^{1}$
          \and
          C. P. Guti\'errez$^{1,2,3}$
          \and
          L. Dessart$^{4}$
          \and 
          M. Hamuy$^{3,2}$
          \and
          L. Galbany$^{2,3}$
          \and
          N. I. Morrell$^{5}$
          \and
          M. D. Stritzinger$^{6}$
          \and
          M. M. Phillips$^{5}$
          \and
          G. Folatelli$^{7,8}$
          \and
          H. M. J. Boffin$^{1}$
          \and
          T. de Jaeger$^{2,3}$
          \and
          H. Kuncarayakti$^{2,3}$
          \and
          J.~L. Prieto$^{9,2}$          }

   \institute{$^1$European Southern Observatory, Alonso de C\'ordova 3107, Casilla 19, Santiago, Chile. \email{janderso@eso.org}\\
			  $^2$Millennium Institute of Astrophysics, Casilla 36-D, Santiago, Chile\\
			  $^3$Departamento de Astronom\'ia, Universidad de Chile, Camino El Observatorio 1515, Las Condes, Santiago, Chile\\
              $^4$Laboratoire Lagrange, Universit\'e C\^{o}te d'Azur, Observatoire de la C\^{o}te d'Azur, CNRS, Boulevard de l'Observatoire, CS 34229, 06304 Nice cedex 4, France. \\
              $^5$Carnegie Observatories, Las Campanas Observatory, Casilla 601, La Serena, Chile\\
              $^6$Department of Physics and Astronomy, Aarhus University, Ny Munkegade 120, DK-8000 Aarhus C, Denmark\\
              $^7$Facultad de Ciencias Astron\'omicas y Geofísicas, Universidad Nacional de La Plata (UNLP), Paseo del Bosque S/N, B1900FWA, La Plata, Argentina; Instituto de Astrof\'isica de La Plata, IALP, CCT-CONICET-UNLP, Argentina\\
              $^8$Kavli Institute for the Physics and Mathematics of the Universe (WPI), The University of Tokyo, Kashiwa, Chiba 277-8583, Japan\\
              $^{9}$N\'ucleo de Astronom\'ia de la Facultad de Ingenier\'ia, 
              Universidad Diego Portales, Av. Ejército 441 Santiago, Chile\\}

   \date{}

% \abstract{}{}{}{}{} 
% 5 {} token are mandatory
 
  \abstract
   {Spectral modelling of type II supernova atmospheres indicates a clear dependence 
   of metal line strengths on progenitor metallicity.
   This dependence motivates further work to evaluate the accuracy with which these
   supernovae can be used as environment metallicity indicators.}
   {To assess this accuracy we present a sample of type II supernova
   host \hii-region spectroscopy, from which environment oxygen
   abundances have been derived. These environment abundances are compared to the observed strength of metal 
   lines in supernova spectra.}
   {Combining our sample with measurements from the literature, we present oxygen abundances of 119
   host \hii\ regions by extracting
   emission line fluxes and using abundance diagnostics. 
   These abundances are then compared
   to equivalent widths of \feii\ 5018\,\AA\ at various time and colour epochs.}
   {Our distribution of inferred type II supernova host \hii-region abundances has a range of $\sim$0.6 dex. We confirm
   the dearth of type II supernovae exploding at metallicities lower than those found (on average) in the Large Magellanic Cloud.
   The equivalent width of \feii\ 5018\,\AA\ at 50 days post-explosion shows a statistically significant 
   correlation with host \hii-region oxygen abundance. The strength of this correlation 
   increases if one excludes abundance measurements derived far from supernova explosion sites.  
   The correlation significance also increases if we only analyse a `gold' IIP sample, and
   if a colour epoch is used in place of time. 
   In addition, no evidence is found 
   of a correlation between progenitor metallicity and supernova light-curve or spectral
   properties -- except for that stated above with respect to \feii\ 5018\,\AA\ equivalent widths -- 
   suggesting progenitor metallicity is not a driving factor in producing
   the diversity that is observed in our sample.}
   {This study provides 
   observational evidence of the usefulness of type II supernovae as metallicity indicators. 
   We finish with a
   discussion of the methodology needed to use supernova spectra as independent metallicity
   diagnostics throughout the Universe.}


   \maketitle
%
%________________________________________________________________

\section{Introduction}
A fundamental parameter in our understanding of the evolution of galaxies is the
chemical enrichment of
the Universe as a function of time and environment. Stellar evolution and
its explosive end drive
the processes which enrich the interstellar (and indeed intergalactic) medium
with heavy elements. 
Galaxy formation and evolution, together with the evolution of the 
complete Universe
are controlled by the speed and temporal location of chemical enrichment. This is
observed in the strong correlation between a galaxy's mass and its 
gas-phase oxygen abundance (see \citealt{tre04}). One also
observes significant radial metallicity gradients within galaxies (see e.g.
\citealt{hen99,san14}) which provides clues to their past formation history and future
evolution.\\
\indent To determine the rate of
chemical enrichment as a function of both time and environment,
metallicity indicators throughout the Universe are needed. 
In nearby galaxies one can use spectra of individual stars 
to measure stellar metallicity (see, e.g. \citealt{kud12}). 
However, further afield this becomes impossible and other
methods are required. 
In relatively nearby galaxies ($<$70 Mpc) one can observe the stellar light from clusters
to constrain stellar metallicities (see e.g. \citealt{gaz14}), or 
gas-phase abundances can be obtained through observations of emission lines
within \hii\ regions produced by the ionisation (and subsequent recombination)
of the interstellar medium (ISM) (see e.g. review of various techniques in \citealt{kew08}). 
At higher redshifts, the latter emission line diagnostics become the dominant source of measurements.\\
\indent Emission line diagnostics can be broadly separated into two groups. The first group, so called empirical 
methods, are those where the ratio of strong emission lines within \hii-region spectra
are calibrated against abundance estimations from measurements of the electron temperature ($T_{\rm e}$, 
referred to as a direct method and derived from the ratio of 
faint auroral lines, e.g. \oiii\ 4363\,\AA\ and 5007\,\AA, see \citealt{ost06}). Some of the most popular empirical relations
are those presented in \cite{pet04}, which use the ratio of \ha\ 6563\,\AA\ to \nii\ 6583\,\AA\ (the N2 diagnostic), or
a combination of this with the ratio of \hb\ 4861\,\AA\ to \oiii\ 5007\,\AA\ (the O3N2 diagnostic). These diagnostics
were updated in \cite{mar13} (henceforth M13), and we use the latter for the main 
analysis in this paper.
The second group of diagnostics are those which use the comparison of 
observed emission line ratios with those predicted by photoionisation/stellar population-synthesis models 
(see e.g. \citealt{mcg91,kew02}).
A major issue currently plaguing absolute metallicity determinations is the 
varying results that are obtained with different line diagnostics.
For example, the photoionisation model methods
generally give abundances that are systematically 
higher than those derived
through empirical techniques.
\cite{lop12} published a review of the systematics involved between the various abundance diagnostics.
It should also be noted that the majority of these techniques use oxygen abundance as a proxy for metallicity,
neglecting elemental variations among metals.\\
\indent Given the number of issues with current metallicity diagnostics, 
any new independent technique is
of significant value. 
\cite{des13} presented type II supernova (SN~II) model spectra 
produced from progenitors with 
distinct metallicity. \citeauthor{des14} (2014; hereafter D14) then showed how the strength of metal lines observed within photospheric phase
spectra are strongly dependent on
progenitor metallicity. Here we present spectral observations of 
SNe~II in comparison to abundances inferred from host \hii-region emission line spectra. 
This comparison presents observational evidence that these explosive events 
may indeed be used as metallicity indicators throughout the Universe.\\

\indent SNe~II are the most frequent stellar explosion in the Universe \citep{li11}.
They are the result of massive stars ($>$8-10\msun) that undergo core collapse at the end of their
lives. The type II designation indicates these events have strong hydrogen features
in their spectra (see \citealt{min41}, and \citealt{fil97} for a review of SN spectral classifications), implying 
their progenitors have retained a significant fraction of hydrogen prior to exploding. 
Historically SNe~II have been separated into II-Plateau (IIP), showing an almost constant luminosity
for 2--3 months
in their light-curves post maximum, and II-Linear (IIL) which decline faster in a 
`linear' manner post maximum \citep{bar79}. However, recent large samples have
been published which question this distinction and argue for a continuum in SN decline rates and other properties
(\citealt{and14a}; A14, and \citealt{san15_2}, although see \citealt{far14a,far14b,arc12} for distinct conclusions).
(In the rest of the manuscript we simply refer to all types as `SNe~II', and differentiate events
by specific photometric/spectroscopic parameters where needed.)
It is clear that SNe~II show significant dispersion in their light-curves and
spectral properties (see e.g. A14, \citealt{and14b} and \citealt{gut14}), and A14 and
\cite{gut14} have speculated (following earlier predictions; see \citealt{bli93}) 
that this observed dispersion could be the result of explosions of progenitors with distinct 
hydrogen envelope masses at death.\\
\indent In D14, a conceptual study of SNe~II as environment metallicity indicators was published 
(following \citealt{des13} in which the impact of various stellar and explosion parameters on the resulting
SN radiation was examined).
Model progenitors of increasing metallicity produced spectra with metal-line equivalent widths (EWs)
of increasing strength at a given post-explosion time or colour. This is the result of the fact that the 
hydrogen-rich envelope -- which is the region probed during
the photospheric phase of SN~II evolution -- retains its original composition (given that
nuclear burning during the stars life or the explosion has
negligible/weak influence on the hydrogen-rich envelope metal content).
Hence, the strength of metal line EWs measured during the `plateau' phase of SNe~II is essentially dependent on the 
abundance of heavy elements contained within that part of the SN ejecta, together with the temperature of the line forming region. 
The results of D14 therefore make a prediction that SNe~II with lower metal line EWs
will be found within environments of lower metallicity within their host galaxies. 
The goal of the current paper is to test such predictions by observing
SN~II host \hii\ regions, and compare SN pseudo--EWs (pEWs) to host \hii-region metal abundances.\\
\indent The manuscript is organised in the following way. In the next section the data sample is introduced, both
of the SNe~II, and of host \hii-region spectroscopy. This is followed by a brief description of 
spectral models. In section 3 we summarise the analysis methods, and in section 4
the results from that analysis are presented. In section 5 the implications of these results are discussed, together with 
future directions of this research. Finally, in section 6 we draw our conclusions.

%__________________________________________________________________

\section{Data sample and comparison spectral models}
The data analysed in this publication comprise two distinct types of observations. 
The first is of SN~II optical spectroscopy obtained during their photospheric phases, i.e. 
from discovery to at most $\sim$100 days post explosion. These data are used to extract 
absorption line pEW
measurements. The second data set is emission line spectral 
observations of host \hii\ regions of SNe~II. These are
used to estimate SN~II environment oxygen abundances, which can be used as metallicity proxies.
In the course of this work we compare our observational results with the predictions from the 
spectral models of D14. The details of these models are briefly summarised below.\\

\subsection{Supernova observations}
Our SN sample comprises $>$100 SNe~II observed by the \textit{Carnegie Supernova Project} (CSP, \citealt{ham06}) 
plus previous SN~II follow-up surveys (`CATS et al.' \citealt{gal16}, sources listed in A14). 
A list of SNe~II included in this analysis is given in the Appendix
Table A.1, together with various parameters from A14 and \cite{gut14}.
In Table A.1 we also list the host galaxy properties: recession velocity and absolute $B$-band
magnitude. The mean host galaxy absolute magnitude is --20.5, and the lowest
host magnitude is --17.7. The vast majority of the SN sample have host galaxies intrinsically 
brighter than
the Large Magellanic Cloud (LMC),\footnote{HyperLeda: http://leda.univ-lyon1.fr/} suggesting 
the vast majority of the sample have environment metallicities
higher than those generally found in the LMC (assuming the accepted luminosity--metallicity relation). 
This is important for the discussion presented later with respect to a lack of
SNe~II in low metallicity environments.\\
\indent Optical low-resolution (typical spectral resolutions between 5 and 8\,\AA, FWHM) spectroscopic time series were obtained 
for SNe~II from 
epochs close to explosion out to nebular phases through a number 
of SN follow-up campaigns. We do not go into the details of the follow-up
surveys here, however more information can be found in a number of previous publications
(see e.g. \citealt{ham03,ham06,ham09,con10,fol10}).
Initial analyses of these spectroscopic data focussing on the nature of the dominant
\ha\ line can be found in \cite{gut14} and \cite{and14b}, while the full data release
and analysis will be published in upcoming papers (Guti\'errez et al. in preparation).\\
\indent The data were obtained with a range of instruments in various forms of long
slit spectroscopy. Data reduction was achieved in the standard manner
using routines within \iraf\footnote{\iraf\ is distributed 
by the National Optical Astronomy Observatory, which is operated by the 
Association of Universities for Research in Astronomy (AURA) under 
cooperative agreement with the National Science Foundation.}, including bias-subtraction; flat-field normalisation;
1d spectral extraction and sky-subtraction; and finally, wavelength and flux calibration.
More details of this process as applied to CSP SN~Ia spectroscopy can be found in
\cite{fol13_2}.

\begin{figure}
\includegraphics[width=9cm]{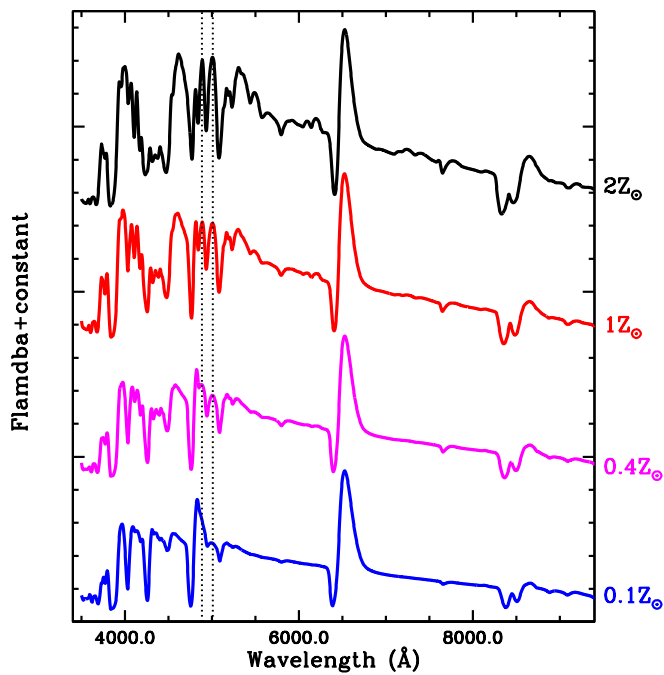}
\caption{Model spectra (D14) from four distinct progenitor models at 50 days post explosion: 0.1, 0.4, 1 and 2
$\times$ solar metallicity. The dotted black lines bracket the \feii\ 5018\,\AA\ absorption feature
we use in this analysis.}
\label{model50}
\end{figure}

\subsection{\hii-region spectroscopy}
In ESO period P94 (October 2014 -- March 2015) 
50 hrs of VLT (+ FORS2) time at Cerro Paranal were allocated to this project.
This was to observe $\sim$100 \hii\ regions coincident or near the site of SNe~II. 
SNe~II were taken from the publications of A14,
plus other SNe~II from the CSP (a small number of `normal' SN~II which were not presented in A14, plus a few IIn and 
IIb), i.e. the same sample
as discussed above with respect to transient optical wavelength spectroscopy.
Measurements from these emission line spectra are also combined with those of other SNe~II which
were previously presented in
\cite{and10} (where many values were taken from \citealt{cov07}). In Appendix Table B.2 
the source of the abundance measurements (here, or from \citealt{and10}) is indicated.\\
\indent SN~II host \hii\ regions were observed using VLT--FORS2 \citep{app98} in long-slit mode (LSS).
We used the 300V grating together with the GG435 blocking filter and a 1\as\ slit. 
This set-up provided a wavelength range of 4450--8650\,\AA, with a resolution of 1.68\,\AA\ pixel$^{-1}$. 
As our target SNe~II are no longer visible (a requirement for our observations and 
analysis methods),
to centre the slit on SN~II explosion sites the telescope was first aligned to a nearby bright star. 
Blind offsets to the SN location were then applied and the slit
position angle was chosen to intersect the SN host-galaxy nucleus.\\
\indent Data reduction was performed in the standard manner using \iraf, in the form of: bias-subtractions;
flat-field normalisations; 1d spectral extraction and sky-subtraction of emission line spectra; and finally wavelength and
flux calibration. One dimensional spectral extraction was first achieved on the exact region
where each SN exploded. However, in many cases no emission lines were detected in that region
(consistent with the non-detection of \ha\ within SN~II environments as reported in \citealt{and12}),
and extractions were attempted further along the slit in either direction until
sufficient lines (at a minimum \ha\ and \nii) could be detected.
The distances of these extraction regions from those of SN explosion coordinates are listed in the Appendix Table~\ref{hiilist},
and the effect of including \hii-region measurements offset from explosion sites is discussed below.\\

\begin{figure*}
\includegraphics[width=9cm]{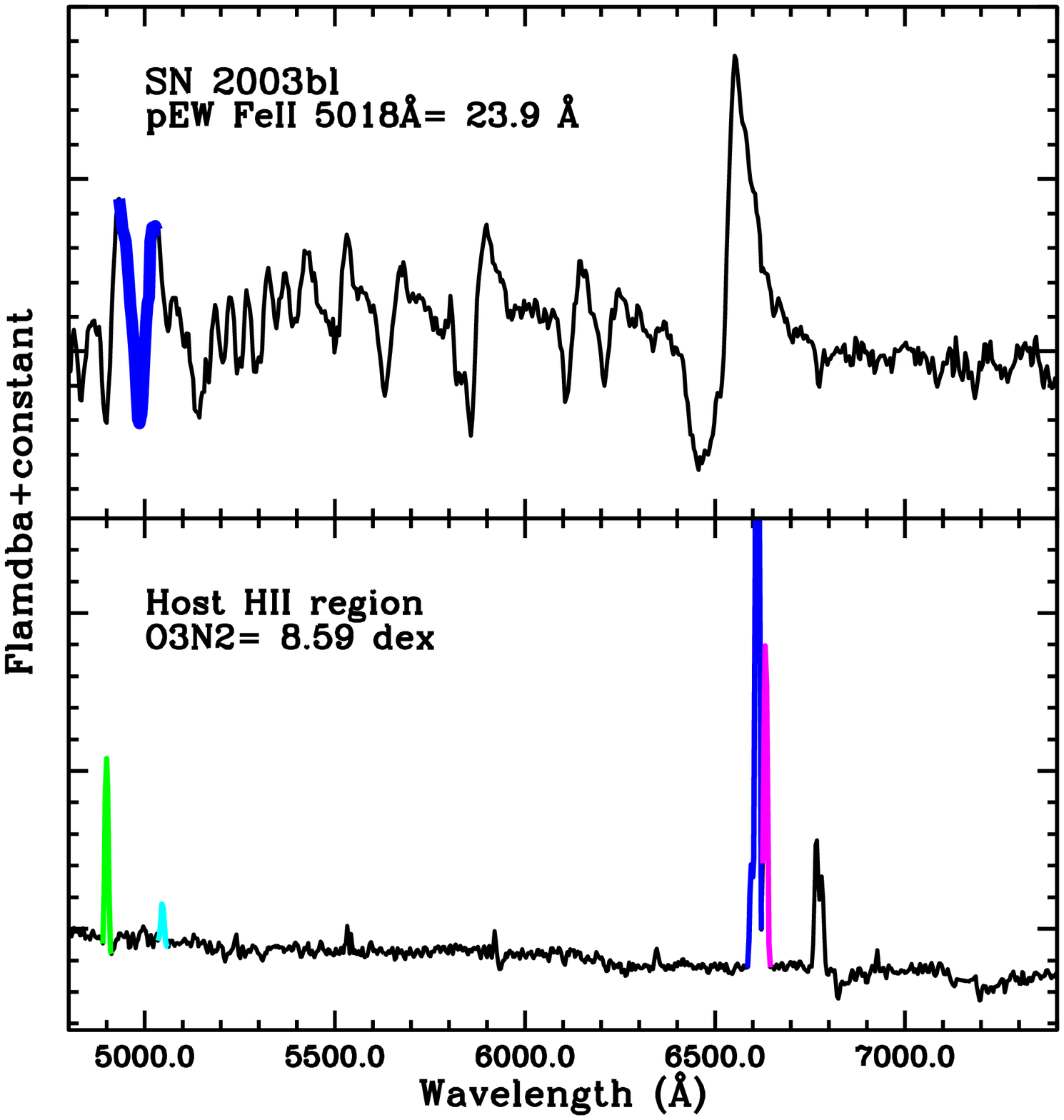}
\includegraphics[width=9cm]{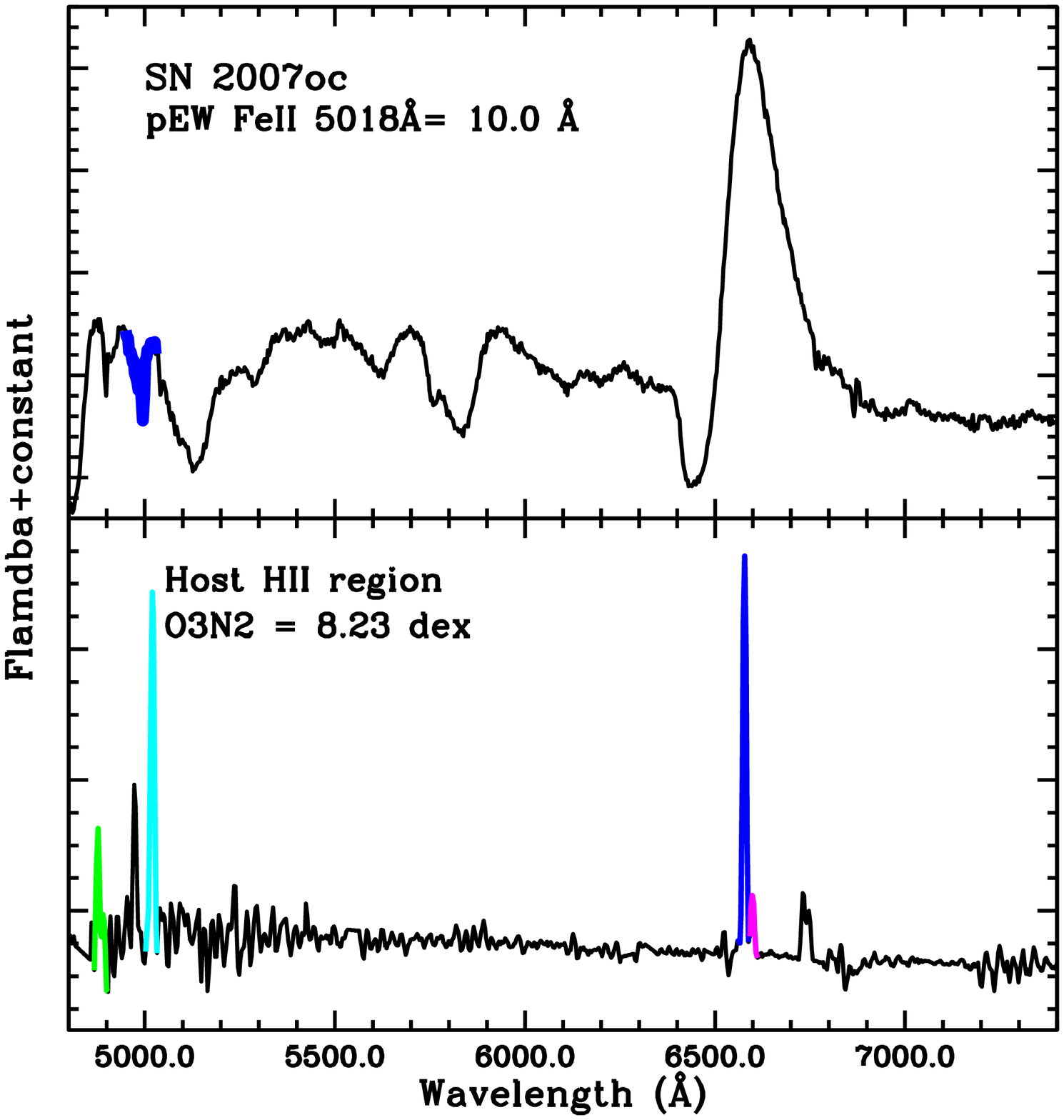}
\caption{Two examples of both SN~II spectra (upper panels, at epochs close
to 50 days post explosion) and host \hii-region spectra (lower panels). \textit{Left:} SN~2003bl, 
a SN with relatively high \feii\ 5018\,\AA\ pEW and O3N2 abundance, and \textit{Right:} SN~2007oc with 
relatively low \feii\ 5018\,\AA\ pEW and O3N2 abundance. The position of the \feii\ 5018\,\AA\ absorption feature
in the SN spectra is indicated in blue. In the \hii-region spectra we indicate the position of the 
emission lines used for abundance estimations: \hb\ in green,
\oiii\ in cyan, \ha\ in blue, and \nii\ in magenta.}
\label{examspec}
\end{figure*}

\subsection{Synthetic spectra of SN~IIP at different metallicities}
\label{desmodels}
The observational research presented in this manuscript was motivated
by the study of D14. That study used four models with distinct progenitor
metallicities, producing synthetic spectral time series. 
Progenitors of 15\msun\ initial mass, and
progenitor metallicities of 0.1, 0.4, 1, and 2 times solar (\zsun) were
evolved from the main sequence until death with \mesa\ \citep{pax11} -- adopting \zsun\ = 0.02.
Upon reaching core-collapse, progenitors
were exploded and synthetic spectral sequences computed using \cmfgen\ \citep{des10,des12}.
The reader is referred
to \cite{des13} and D14 for a detailed explanation of the modelling procedure (\citealt{des13} explore
a large range of progenitor parameters and their subsequent effects on the model SNe~II produced,
while D14 concentrates on the effect of progenitor metallicity).\\
\indent In Figure~\ref{model50} model spectra are plotted, one for each progenitor metallicity, taken at
50 days post explosion (50\,d). One can clearly see the effects of increasing metallicity on the model spectra, in 
particular at bluer wavelengths (i.e. bluewards of \ha, $\lesssim$6000\,\AA). The higher metallicity models exhibit many more
lines which are also significantly stronger in EW. As one goes to the lower
metallicity models spectra appear much `cleaner' being dominated by Balmer lines and
showing weaker signs of metal line blanketing.\\ 
\indent D14 explored the effects of changing progenitor
metallicity with all other parameters constant (initial mass, mass-loss and
mixing length prescriptions). However, there are other pre-SN
parameters which may significantly affect the evolution and strength
of spectral line EWs (the important features we use in the current work), and produce
degeneracies in SN measurements.
D14 showed how SNe~II with distinct pre-SN radii (created using the same progenitors,
but evolved with a distinct mixing length prescription for convection) produced different
EW strengths and evolutions for the same progenitor metallicity. This is seen in Figs.~\ref{ewtimebin} and \ref{n2models}, 
where
the models are referred to as m15mlt1 (larger radius) and m15mlt3 
(smaller radius, m15mlt2 is the solar metallicity
model already discussed above).
We use such models to compare the metal line EWs resulting 
from metal abundance variations or from changes in the progenitor structure.

%__________________________________________________________________

\section{Analysis}
As outlined above, our data comprises two distinct sets, and hence
our analysis is split into two distinct types of measurements. These
are now outlined in more detail. In Fig.~\ref{examspec} we present examples of our data, indicating the position of the spectral 
lines used in our analysis.

\subsection{SNe~II \feii\ 5018\,\AA\ EW measurements}
To quantify the influence of progenitor metallicity on 
observed line strengths, in this publication we concentrate on the strength 
of the \feii\ 5018\,\AA\ line. This line is prominent in the majority of SNe~II from relatively early times, i.e. 
at the onset of hydrogen recombination,
and stays present throughout the photospheric phase. In addition, it is not significantly
contaminated by other SN lines. One issue with this line is that it is in the 
wavelength range where one observes narrow \hb\ and \oiii\ emission lines
from host \hii\ regions. Often it is difficult to fully remove these
features
in spectral reduction and extraction, and they can contaminate the broad spectral 
features of the SN. 
When narrow \hii-region emission lines are present in our
SN spectra, they are removed by simply interpolating the SN spectra between either side of the emission line. 
The uncertainty created by this
process is taken into account when estimating flux errors. 
pEWs are measured in all spectra obtained within 0--100 days post explosion (see A14 for details of explosion
epoch estimations).
To measure pEWs we proceed to define the pseudo continuum (the adjacent maxima that bound the
absorption)
either side of the broad SN absorption 
feature and fit a Gaussian. These are defined as `pseudo' EWs due to the difficulty
in knowing/defining the true continuum level. This procedure is achieved multiple times, each time
removing narrow emission lines when present. A mean
pEW is then calculated together with a standard deviation, with the latter being taken
as the pEW error. In this way we obtain a pEW for each spectral epoch.
The same measurement procedure undertaken for observations 
is achieved for model spectra, meaning that for models we also present pEWs to make consistent
comparisons with observations (even though in the case of models we know the true continuum and 
could measure true EWs).

\subsection{Host \hii-region abundance measurements}
\begin{figure}
\includegraphics[width=9cm]{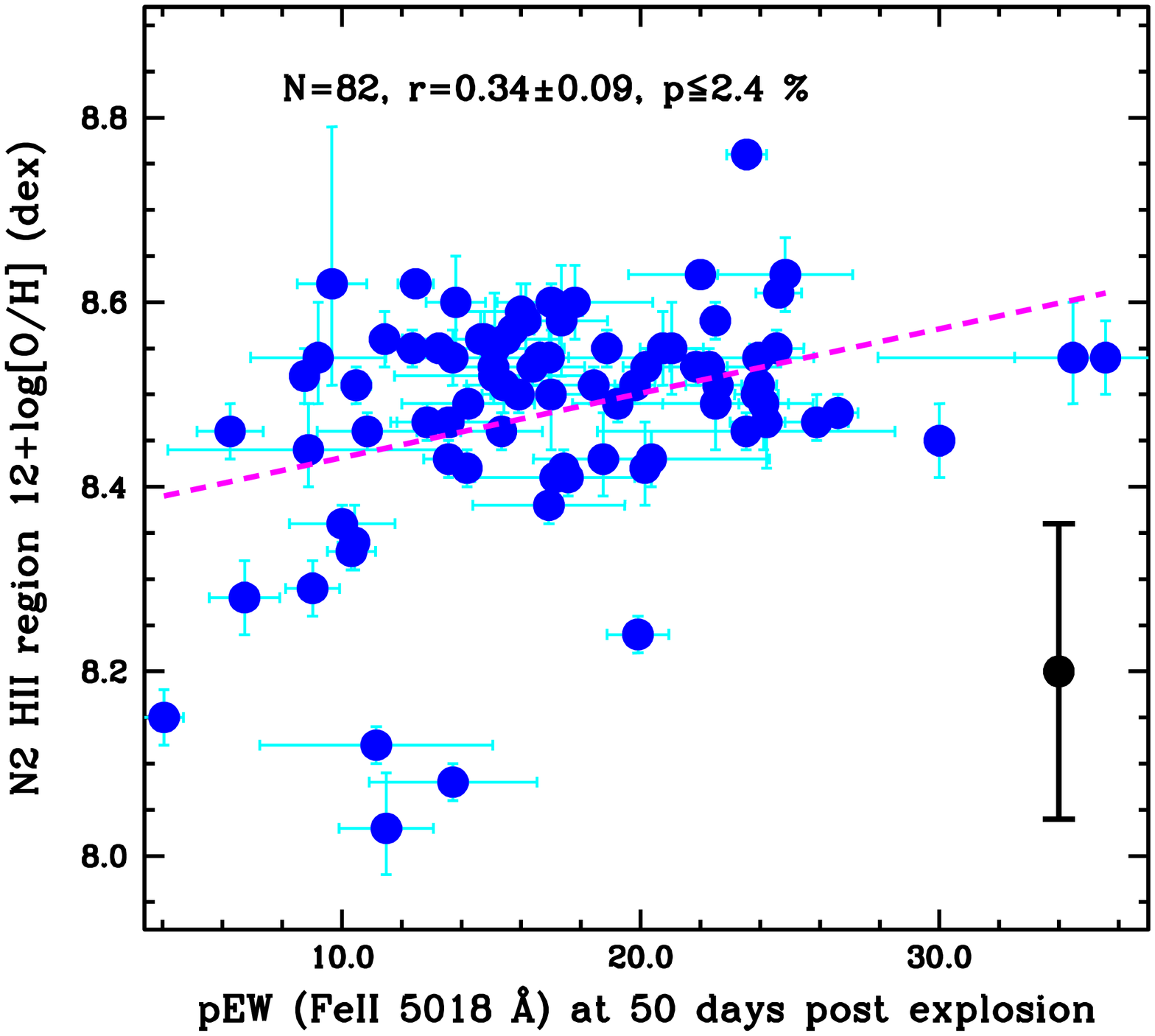}
\caption{pEW of the \feii\ 5018\,\AA\ absorption line measured at 50\,d 
plotted against host \hii-region oxygen abundance on the N2 M13 scale.
The dashed line indicates the mean best fit to the data. Error bars on individual measurements
are the statistical errors from line flux measurements.
The large black error bar gives the N2 diagnostic error from M13.}
\label{ew50_all}
\end{figure}

Fluxes of all detected narrow emission lines within host \hii-region spectra are measured by defining the continuum on either side of the emission and fitting
a Gaussian to the line. The lines of interest for our abundance estimations are:
\hb, \oiii, \ha, and \nii. Using these fluxes, gas-phase oxygen abundances are calculated using the 
N2 and O3N2 diagnostics of both M13 and the earlier calibrations from \cite{pet04}.
These are listed in Table~\ref{hiilist}. 
Abundance errors are estimated by calculating the minimum and maximum line ratios taking 
into account line flux errors, i.e. the `analytic' approach outlined in \cite{bia15} (and used
on the previous sample of \citealt{and10}). In addition, in Figs.~\ref{ew50_all} and ~\ref{O3N2all} the
systematic errors from the M13 N2 and O3N2 diagnostics are also shown.\\
\indent We are restricted in the abundance diagnostics we can use simply because of 
small number of detected emission lines in our data.
Our exposure times were relatively short so in many cases only \ha\ and \nii\ are detected, making the
N2 diagnostic the only possibility. An advantage of both the N2 and O3N2 diagnostics is that they are essentially unaffected
by either host-galaxy extinction and/or relative flux calibration, due to the use of ratios of emission lines
close in wavelength. While both the M13 and \cite{pet04} abundances are listed in Table~\ref{hiilist}, we use the M13
values for our analysis given the recalibration of the diagnostics including additional \hii-region $T_{\rm e}$ 
measurements\footnote{If we were to use the \cite{pet04} values instead then our results and conclusions
remain unchanged.}.

\begin{figure}
\includegraphics[width=9cm]{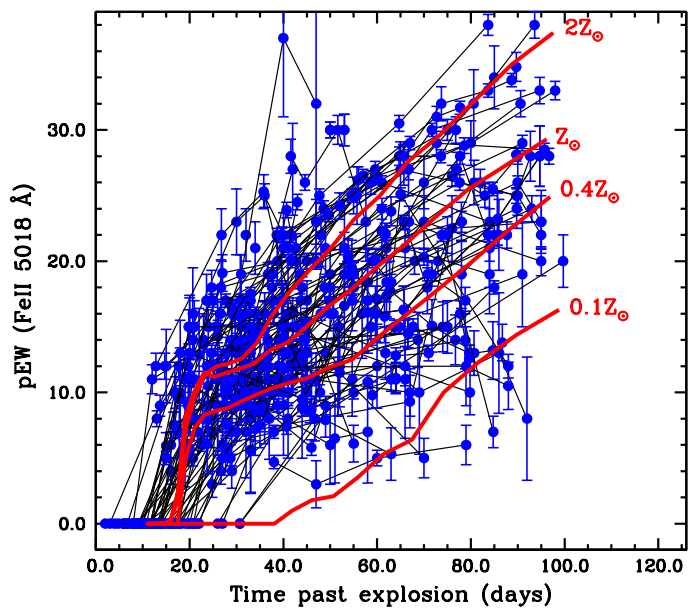}
\caption{Evolution of \feii\ 5018\,\AA\ pEWs with time for all SNe~II within the sample. Individual measurements
are shown in blue, together with their errors. These are connected by black lines. Also presented are the 
time sequence of pEWs measured from synthetic spectra \citep{des13}, for four models of 
distinct progenitor metallicity.}
\label{ewtime}
\end{figure}

\begin{figure}
\includegraphics[width=9cm]{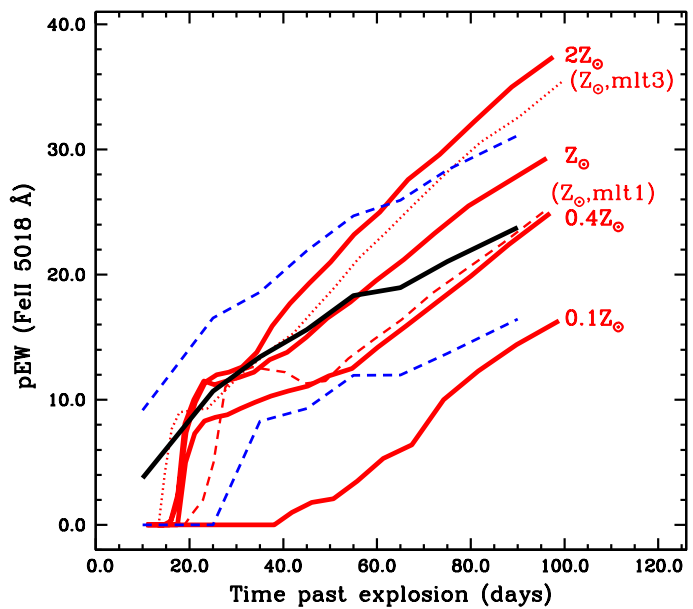}
\caption{Same as Fig.~\ref{ewtime}, but now the observational data are binned in time. The black
sold line represents the mean pEW within each time bin, while the dashed blue lines indicate
the standard deviation. 
Together with the four distinct metallicity models (\citealt{des13} shown in solid red lines)
we also present spectral series produced from two additional solar metallicity progenitors,
but with distinct mixing length prescriptions, leading to smaller (mlt3, shown as the dotted red line) and larger
(mlt1, shown as the dashed red line) RSG progenitor radii (see Section ~\ref{desmodels} for more details).}
\label{ewtimebin}
\end{figure}

\begin{figure*}
\centering
\includegraphics[width=14cm]{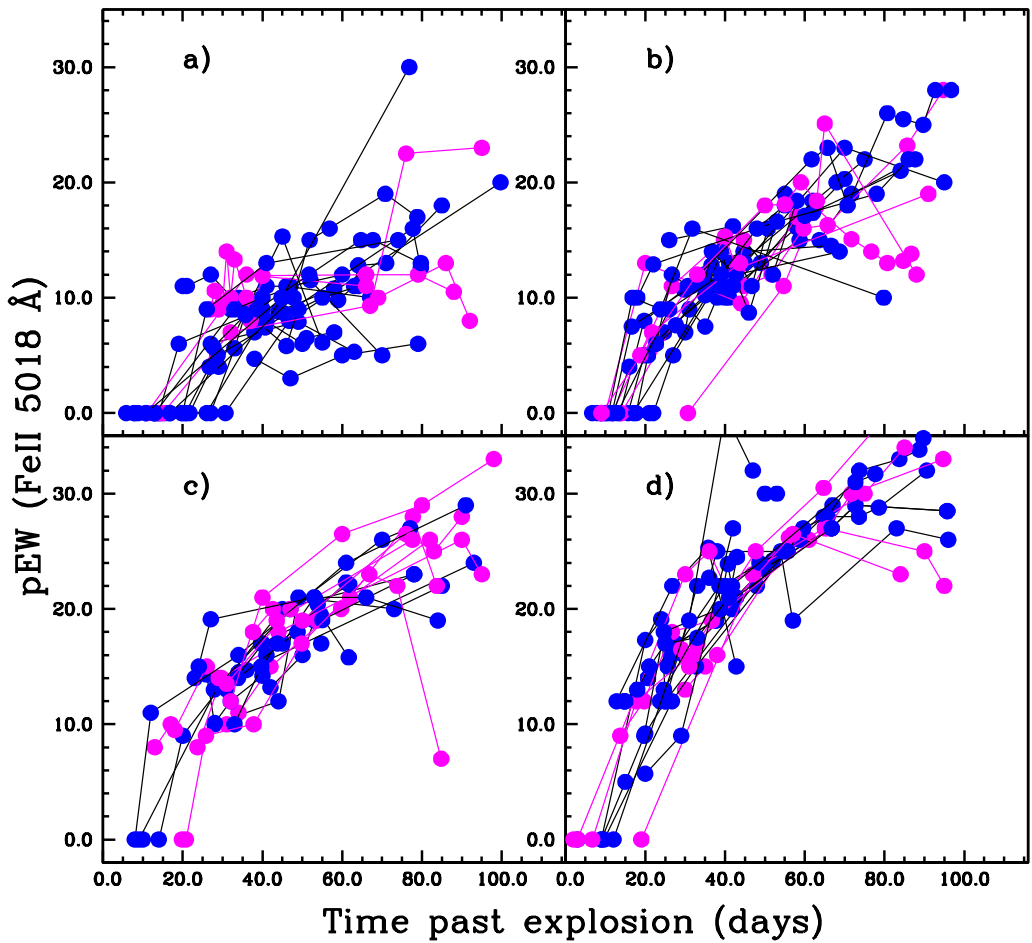}
\caption{Evolution of \feii\ 5018\,\AA\ pEWs with time, as
presented in Fig.~\ref{ewtime}, separated into 4 panels splitting the sample
by their \feii\ 5018\,\AA\ pEWs at 50\,d. The distributions go from low pEW in
panel a) through to the highest pEW sample in panel d). The overall sample
is presented in blue, while the `Gold' sample discussed later is plotted
in magenta. (Errors are not
plotted to enable better visualisation of the trends.)}
\label{ew4pan}
\end{figure*}

\begin{figure}
\includegraphics[width=9cm]{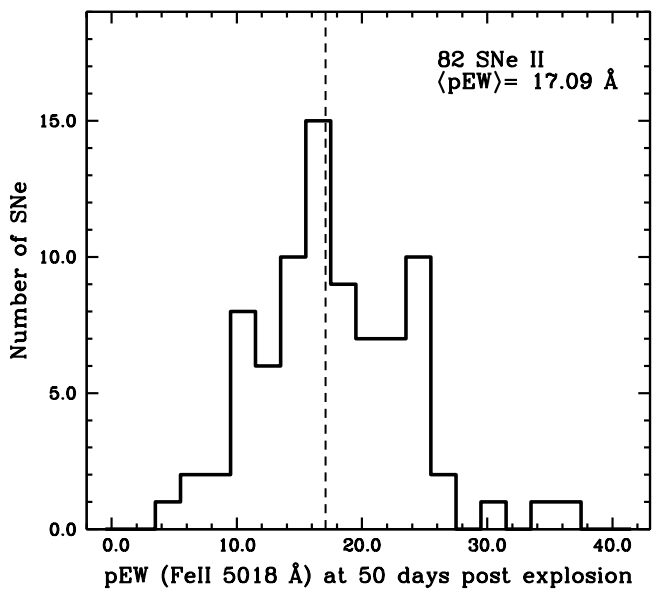}
\caption{Histogram of \feii\ 5018\,\AA\ pEW measurements for our sample of SN~II spectra, 
interpolated to 50\,d. The position of the mean pEW is indicated by the vertical dashed line.}
\label{histEW}
\end{figure}

\begin{figure}
\includegraphics[width=9cm]{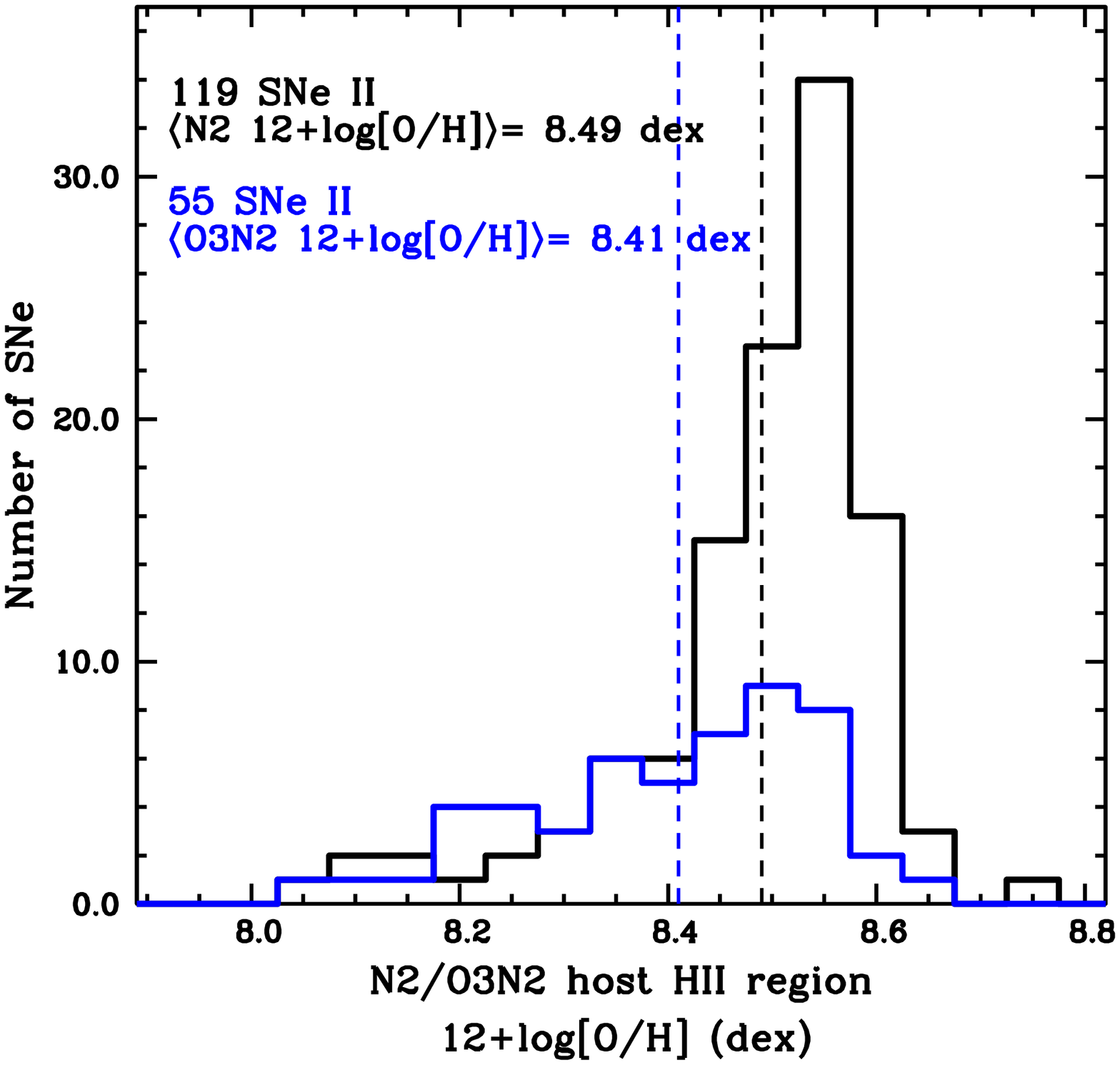}
\caption{M13 N2 and O3N2 abundance
distributions, together with their mean values (dashed lines).}
\label{histZ}
\end{figure}

%__________________________________________________________________

\section{Results}
Above we have presented two sets of observations: spectral line \feii\ 5018\,\AA\ pEW measurements during the
photospheric phase of SNe~II, and emission line spectral measurements of SN~II host 
\hii\ regions, with the latter being used to obtain environment oxygen abundances. 
The distributions of these are now both presented. Then we proceed to correlate both parameters, and
confront model predictions with SN and host \hii-region observations.
In addition, we analyse how pEWs are related to other SN~II light-curve and spectral parameters, 
and finally we search for correlations between environment metallicity and SN~II transient properties.\\

\subsection{\feii\ 5018\,\AA\ pEW distribution and evolution}
The time evolution of SN pEWs is shown in Fig.~\ref{ewtime}, together with those from spectral
models. In Fig.~\ref{ewtimebin} we present these same measurements but now with
the observations binned (with bins of 0--20 days, then 20--30, 30--40, 40--50, 50--60, 60--70, 70--80, and
80--100 days). 
In both figures the evolution of model pEWs is also presented.
These figures show the increase in time of pEWs (using the convention that a deeper absorption
is documented as a larger positive pEW), but also the large dispersion between different
SNe~II.  
One can also see the distinct strength and evolution of \feii\ 5018\,\AA\ pEWs
found within model spectra from the four distinct progenitor metallicities.
The effect on model spectra of changing pre-SN radii (at a fixed, solar, metallicity)
can be seen in Fig.~\ref{ewtimebin} (m15mlt1 larger radius, and m15mlt3 
smaller radius).\\
\indent In Fig.~\ref{ew4pan} the pEW -- time past explosion trends
from Fig.~\ref{ewtime} are again presented but now split into 4 panels,
separating SNe by their pEWs at 50\,d post explosion (see below for discussion
of this measurement). It is now possible to observe the trends of individual
SNe in more detail. This confirms the monotonic behaviour of
\feii\ 5018\,\AA\ pEWs
with time past explosion throughout the `plateau' phase of SNe~II evolution
($<$100 days post explosion), in qualitative agreement with models. The plot
also shows that SNe with similar pEWs at 50\,d evolve in a similar manner with
relatively
low dispersion.\\
\indent To proceed with our analysis pEWs are required at consistent epochs between SNe. 
Our epoch of choice is 50\,d (more details below). To estimate the pEW at this epoch,
interpolation/extrapolation is needed. This is achieved for SNe~II with $\geq$2 measurements available
and where in the case of extrapolation, a spectrum is available at 50$\pm$10 days post explosion. 
Then a low order polynomial fit is made to the pEW measurements, and this is used 
to obtain a pEW at 50\,d. We use the RMS error of this fit as the error on the interpolated
pEW.
This interpolation was possible in 82 cases, and 
a histogram of these measurements is presented in Fig.~\ref{histEW}.

\subsection{Host \hii-region abundances}
In Fig.~\ref{histZ} we present the distribution of all SN~II emission line
abundance measurements for both the N2 and O3N2 M13 diagnostics. 
The N2 distribution has a mean value of 12+log[O/H] = 8.49 dex,
and a median of 8.52 dex. 
The distribution
shows a peak at just below $\sim$8.6 dex, and a tail out to lower abundances
with the lowest value of 8.03 dex.
The O3N2 distribution has a mean of 8.41 dex, and a median value of 8.44 dex. 
The distribution
shows a peak at $\sim$8.5 dex, a range of $\sim$0.6 dex, and a tail out to lower abundances
with the lowest value of 8.06 dex.
Using a solar value of 12+log[O/H] = 8.69 dex \citep{asp09}, the O3N2 distribution thus
ranges between 0.23 and 0.87\zsun, with a mean of 0.51\zsun. However, we stress that
any discussion of absolute metallicity scale when dealing with emission line diagnostics
is problematic, and it is probable that the N2 and O3N2 diagnostics give systematically lower
abundances than the true intrinsic values (see e.g. \citealt{lop12} and references therein).\\

\subsection{SN~II \feii\ 5018\,\AA\ pEWs and host \hii\ region abundance}
In Fig.~\ref{ew50_all} SN~II \feii\ 5018\,\AA\ pEWs at 50\,d are plotted against the N2 diagnostic
on the M13 scale. 
To test the significance of this trend, and all subsequent correlations, we run a  
Monte Carlo simulation randomly selecting events from the distribution in a bootstrap 
with replacements manner 10000 times. 
The mean Pearson's correlation $r$ value
is determined together with its standard deviation using the 10000 random sets
of pEW -- abundances. The lower limit of the 
chance probability of finding a correlation,
$p$, is then inferred using these values\footnote{It is generally
considered that for $r$ values between 0.0 and 0.2 there is zero or negligible correlation; between 0.2 and 0.3 
weak correlation; between 0.3 and 0.5 moderate correlation; and above 0.5 signifies strong correlation. The $p$ value
gives the probability that this level of correlation is found by chance.}.
With a total of 82 events which have SN~II \feii\ 5018\,\AA\ pEWs and N2 oxygen abundance measurements,
we find $r$ = 0.34$\pm$0.09, and a chance
probability of finding a correlation of $\leq$2.4 \%. Statistics using only the observed distribution
give: $r$ = 0.34, $p$ = 0.18\%. Hence, using N2 we find a moderate strength correlation between 
SN pEW and host \hii-region abundance. It is clearly observed that there is a lack of SNe~II with high pEW
and low abundance at the bottom right of Fig.~\ref{ew50_all}.
There are fewer SN environments where we were able to also measure (in addition to \ha\ and \nii) the 
\hb\ and \oiii\ fluxes needed to compute abundances on the O3N2 scale. 
We are able to do this in 44 cases, and here we obtain a mean $r$ value of 0.50$\pm$0.10,
which gives a chance probability of $\leq$0.7 \%. Statistics using only the observed distribution
give: $r$ = 0.50, $p$ = 0.05 \%. The correlation is shown in Fig.~\ref{O3N2all}. While the statistical
significance is higher for the O3N2 diagnostic (and therefore we use that diagnostic for subsequent 
sub-samples), both of these figures show there
is a statistically significant trend in the direction predicted by models:
SNe~II with larger \feii\ 5018\,\AA\ pEWs tend to be found in environments of higher oxygen abundance.
These observational results hence agree with model predictions, and motivate further work
to use SNe~II as environment metallicity indicators. We also note that the
RMS errors on the M13 N2 and O3N2 diagnostics (as plotted in Figs.~\ref{ew50_all} and ~\ref{O3N2all}), appear
to be large when compared to the spread of values in our plots. Given that we do find evidence for correlation, this
suggests the true precision of those diagnostics for predicting \hii-region abundance is better
than the values given by M13.\\
\indent In Fig.~\ref{n2models} we over-plot SN~II models at four different metallicities,
together with the additional two models at solar metallicity but with different pre-SN radii (D14), onto the 
observed \feii\ 5018\,\AA\ pEW vs.\ \hii-region abundance plot. The model metallicities are
converted from fractional solar to oxygen abundance using a solar value of 8.69 dex \citep{asp09}. 
One can see that the models produce a much steeper trend than
that observed. We also see that models with the same progenitor metallicity but with
different pre-SN radii produce a range of almost 10\,\AA\ in \feii\ 5018\,\AA\ pEWs.
This uncertainty can be reduced with a slight time shift (of pEW measurements)
or by comparing at a given colour (see below for additional analysis).
However, in general, SN~II observations favour 
relatively low progenitor radii \citep{des13,gon15}, and explosions similar to that of the
large radius model are probably rare in nature. 
Therefore the actual uncertainty in observed SNe~II is 
probably less than that represented by this range of models.
Another interesting observation from Fig.~\ref{n2models} is the lack of any SN close to the tenth
solar model. There is also a lack of SNe~II at super-solar values. We discuss this
(possibly small) range of metallicities probed by observations below.

\subsubsection{Sub-samples}
\begin{figure*}
\centering
\includegraphics[width=13cm]{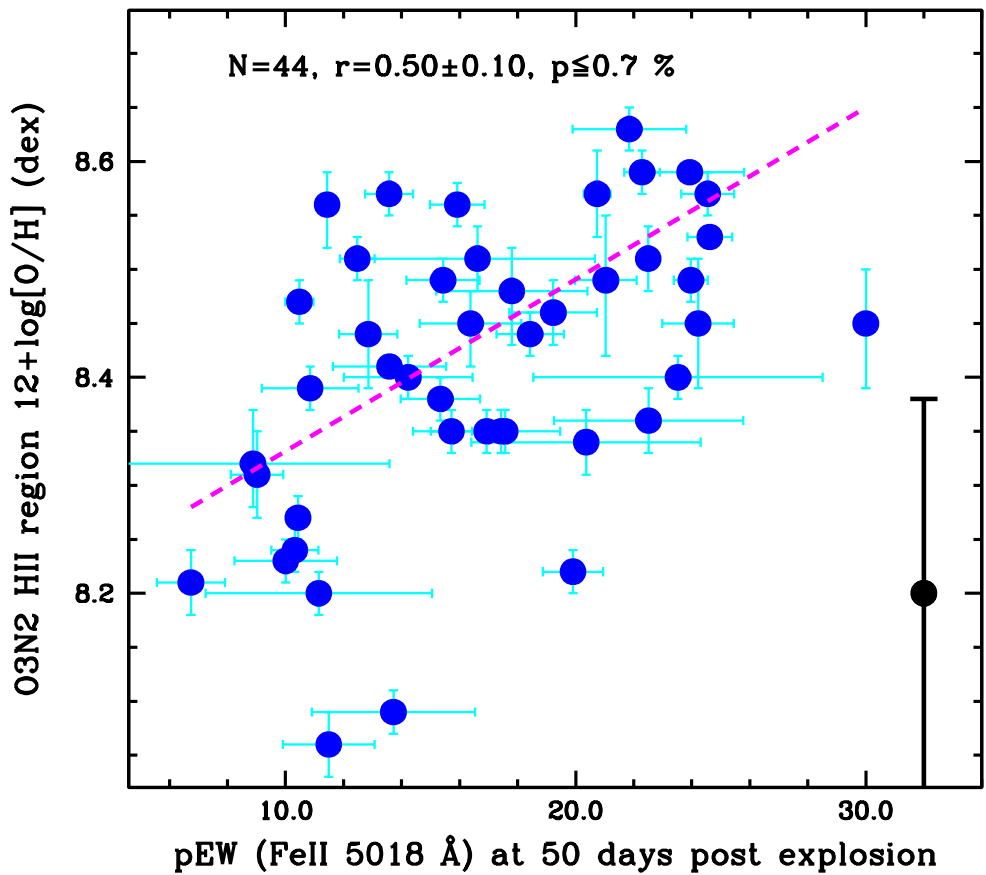}
\caption{Same as Fig.~\ref{ew50_all}, but now for the O3N2 diagnostic on the M13 scale. 
The dashed line indicates the mean best fit to the data. Error bars on individual measurements
are the statistical errors from line flux measurements.
The large black error bar gives the O3N2 diagnostic error from M13.}
\label{O3N2all}
\end{figure*}

\begin{figure}
\includegraphics[width=9cm]{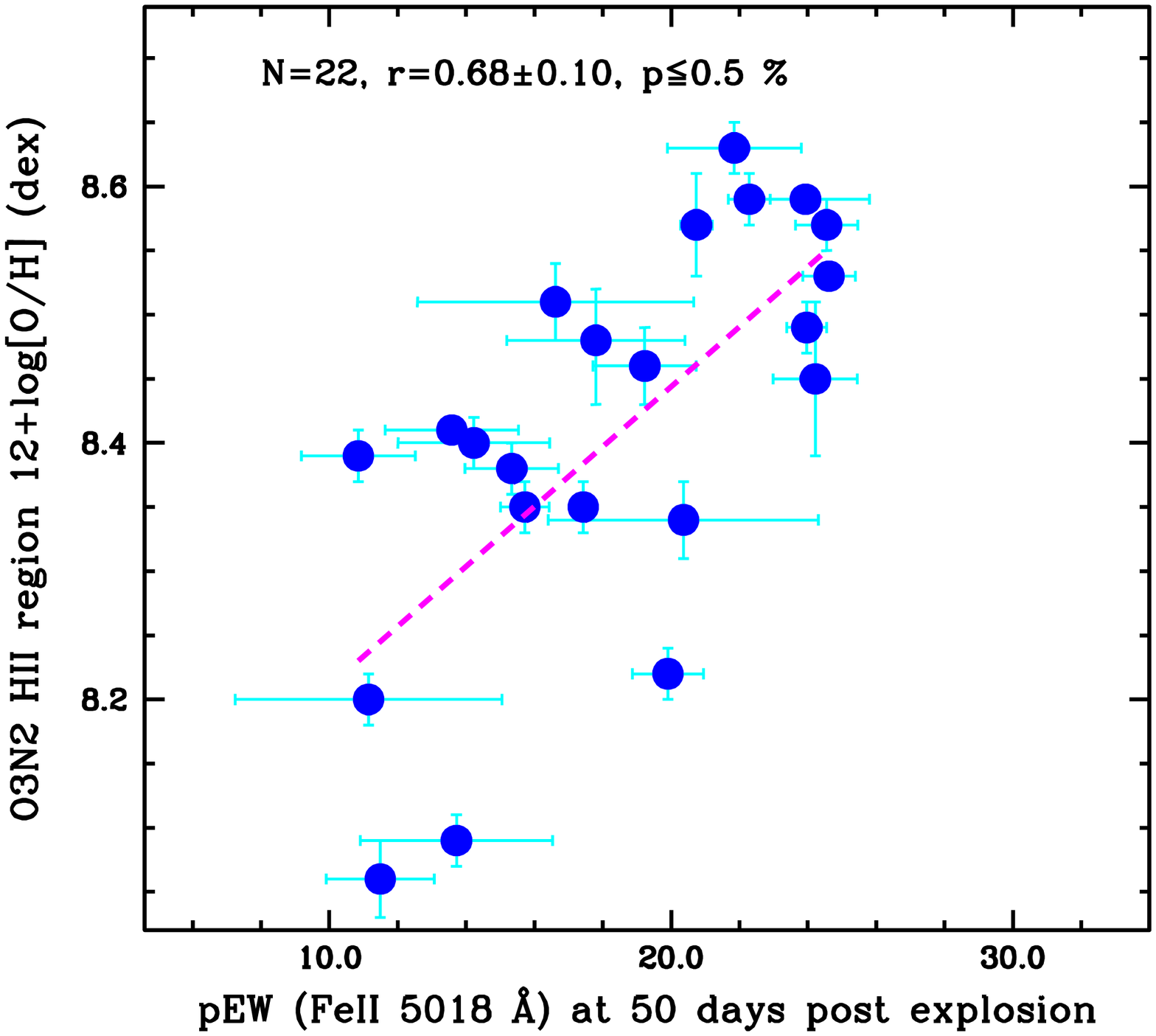}
\caption{`Gold IIP' sample of pEWs against O3N2 abundances, using SNe~II with $s_2$ values 
$\leq$ 1.5 mag per 100 days,  and/or OPTd values $\geq$ 70 days.}
\label{goldII}
\end{figure}

In Figs.~\ref{ew50_all} and ~\ref{O3N2all} all SNe~II were included irrespective of 
their light-curve or spectral properties. This means that we include SNe~II with a wide range of
absolute magnitudes, `plateau' decline rates (s$_2$), and optically thick phase durations (OPTd)
(and other SN parameters which differ from one event to the next). The models of D14 all produce
SN light-curves and spectra typical of `normal' SNe~IIP. In addition,
those figures included all SN~II abundance measurements, including events with abundances estimated from
a region of their host galaxies at significant distances from explosion sites. Here these issues are
further investigated.\\
\indent First we construct a sub-sample of events where abundance estimations were carried out 
less than 2 kpc away from the explosion sites\footnote{This limit is somewhat arbitrary, however it
removes cases where extractions are at a significant distance from explosion sites, while 
maintaining a sufficient number of events to enable a statistically significant analysis.}.
When this is achieved we are left with 56 SNe~II with measurements on the N2 scale and 32 on 
the O3N2 scale. We again test for correlation using Monte Carlo bootstrapping with replacements, and
in the case of N2 a correlation coefficient $r$ of 0.42$\pm$0.10 is found, giving a chance 
probability (N=56) of $\leq$ 1.6 \%. Using O3N2 we obtain $r$ = 0.55$\pm$0.10 and a $p$ value of $\leq$1.0 \%\ (N=32).
The level of correlation thus increases when we only include abundance measurements closer to SN explosion
sites. This is to be expected, as SNe~II have relatively short lifetimes and are therefore not
expected to move significantly from their birth sites. As one moves away from exact explosion sites
environment metallicity becomes less representative of progenitor abundance due to spatial metallicity
changes from one region of a galaxy to another, especially in terms of increasing/decreasing 
galacto-centric offset.\\
\indent To produce a sub-sample of `normal' SNe~IIP cuts are made to our sample in terms
of light-curve morphologies. SNe~II with $s_2$ values $\geq$ 1.5 mag per 100 days,  and/or OPTd values $\leq$ 70 days
are removed from the sample. Using these cuts a `Gold IIP' sample of 22 SNe~II is formed, which would generally be considered
typical SNe~IIP by the community.
Testing for correlation, this sub-sample has an $r$ value = 0.68$\pm$0.10 and a $p$ value = $\leq$0.5 \%.
This correlation is presented in Fig.~\ref{goldII}, and shows the increase in strength of correlation
as compared to the full O3N2 sample in Fig.~\ref{O3N2all} (characterised by
$r$ = 0.50$\pm$0.10). The fast declining ($s_2$ $\geq$ 1.5 mag per 100 days)
sample presents a lower level of correlation (than that of the `normal'
SNe~IIP): 
for 15 SNe, $r$ = 0.57$\pm$0.18 and $p$ = $\leq$4.6 \%. Both the `Gold IIP'
and fast-declining samples show the same linear trends within their
errors. This suggests that `normal' SNe~IIP are better metallicity indicators
than their faster declining counterparts. 

\begin{table*}
\centering
\caption{Epoch of pEW measurements: testing the level of correlation between \feii\ 5018\,\AA\ pEWs and
host \hii-region abundances when different epochs for pEW measurements are used.
In the first column the epoch is listed, followed by the number of
SNe~II available in column 2. The Pearson's mean $r$ value is then given together
with the standard deviation. Finally in column 4 the chance probability of finding
a correlation is listed.}
\begin{tabular}[t]{cccc}
\hline
Epoch & N SNe~II & $r$ & $p$\\
\hline
Days after explosion epoch &&&\\
30 & 90 & 0.30$\pm$0.09 & $\leq$ 4.7 \%\ \\
40 & 83 & 0.32$\pm$0.09 & $\leq$ 3.6 \%\ \\
50 & 82 & 0.35$\pm$0.09 & $\leq$ 1.8 \%\ \\
60 & 75 & 0.30$\pm$0.11 & $\leq$ 10.3 \%\ \\
70 & 68 & 0.38$\pm$0.10 & $\leq$ 2.1 \%\ \\
\hline
With respect to $t_{\rm tran}$ &&&\\
+20 & 39 & 0.24$\pm$0.16 & $\leq$ 63 \%\ \\
Same sample but at 50\,d&&&\\
50 & 39 & 0.30$\pm$0.15 & $\leq$ 36 \%\ \\
\hline
With respect to OPTd &&&\\
--30 & 55 & 0.34$\pm$0.11 & $\leq$ 9.1 \%\ \\
Same sample but at 50\,d&&&\\
50 & 55 & 0.35$\pm$0.11 & $\leq$ 7.8 \\
\hline	
\hline
\end{tabular}
\label{epochtab}
\end{table*}

\subsubsection{The epoch of pEW measurements}
Above we presented the pEW distribution and then correlations with \hii-region abundances using \feii\ 5018\,\AA\ pEWs at 
50\,d. This epoch was chosen as it corresponds to when the vast majority of SNe~II
are around halfway through the photospheric phase of their evolution.
SNe~II can show significant temperature variations at early times and one does not want 
to measure pEWs when temperature differences could be a significant factor controlling their strength.
Fig.~\ref{ewtimebin} also shows that pEWs have a rapid increase at early times after first appearing at
$\sim$15 days, and therefore one wants to avoid this region where non-metallicity systematics may dominate differences
in pEWs. 
At much later than 50\,d some SNe~II already start to transition from the photospheric phase
to radioactively powered epochs. Here, pEWs may start to be affected
by mixing of He-core material (in addition to the fact that spectral observations become more sparse). 
We now test whether this selected epoch is the most appropriate
to measure pEWs.\\
\indent pEW measurements are interpolated to: 30, 40, 60 and 70 days post explosion. 
These values are then correlated against host \hii-region
abundances on the N2 scale (used to ensure sufficient statistics for 
valid comparisons) and the strength of the correlations are compared. In addition, in place of the
explosion epoch, pEWs are estimated with respect to $t_{\rm tran}$: the epoch of transition from 
the initial s$_1$ decline to the slower `plateau' s$_2$ phase (see A14 for details of those measurements).
We choose the epoch $t_{\rm tran}$+20 days, which roughly coincides with 50\,d, but varies (approximately $\pm$10--15 days) between SNe~II. The results of these comparisons are presented in Table~\ref{epochtab}.
In terms of time post explosion we see that the choice of 50 days is in fact valid. Its is also
observed that 50\,d does just as well as $t_{\rm tran}$+20 days.
A time epoch with respect to OPTd is also investigated (OPTd is the OPtically Thick phase time duration, from explosion to the end
of the `plateau', A14). pEWs are interpolated to OPTd--30 days (again to coincide on
average with t = $\sim$50\,d), and we run our correlation tests for both this new OPTd sample, and the same 
SNe~II but with pEWs
at 50\,d. Values are again displayed in Table~\ref{epochtab}, and it is found that using OPTd as the time epoch
is no better than using the explosion epoch.
In conclusion
the choice of 50\,d for pEW measurements appears to be robust.\\ 

\begin{figure}
\includegraphics[width=9cm]{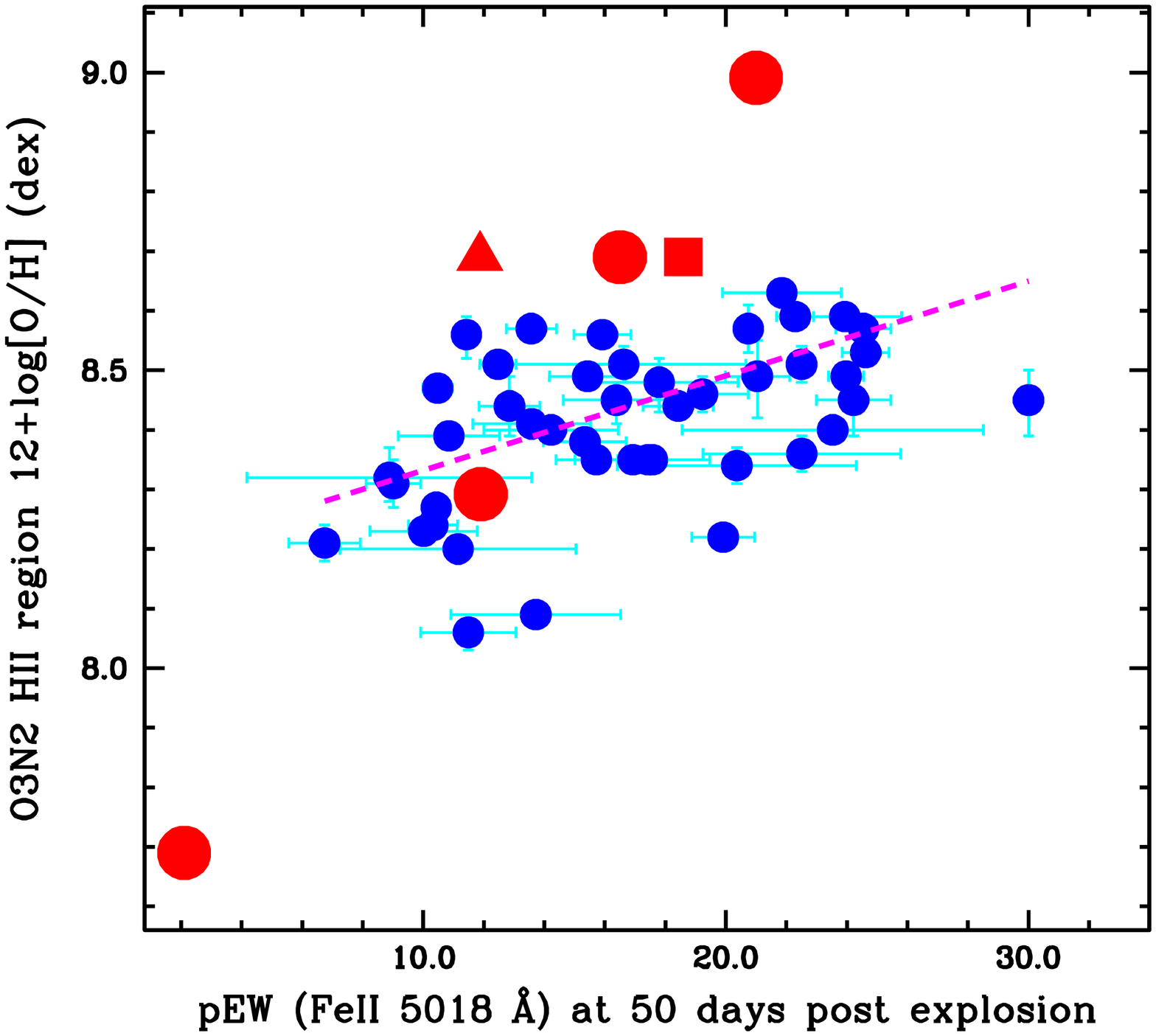}
\caption{Comparison of the D14 models with observations.
pEWs of the \feii\ 5018\,\AA\ absorption line measured at 50\,d 
plotted against the host \hii-region oxygen abundance using the O3N2 M13
diagnostic, with the positions of the six distinct models over plotted. The red circles indicate
the same progenitors changing metallicity, while the triangle (m15mlt1, larger radius) and square (m15mlt3, 
smaller radius) present 
the same metallicity but changing pre-SN radii.}
\label{n2models}
\end{figure}

\begin{figure}
\includegraphics[width=9cm]{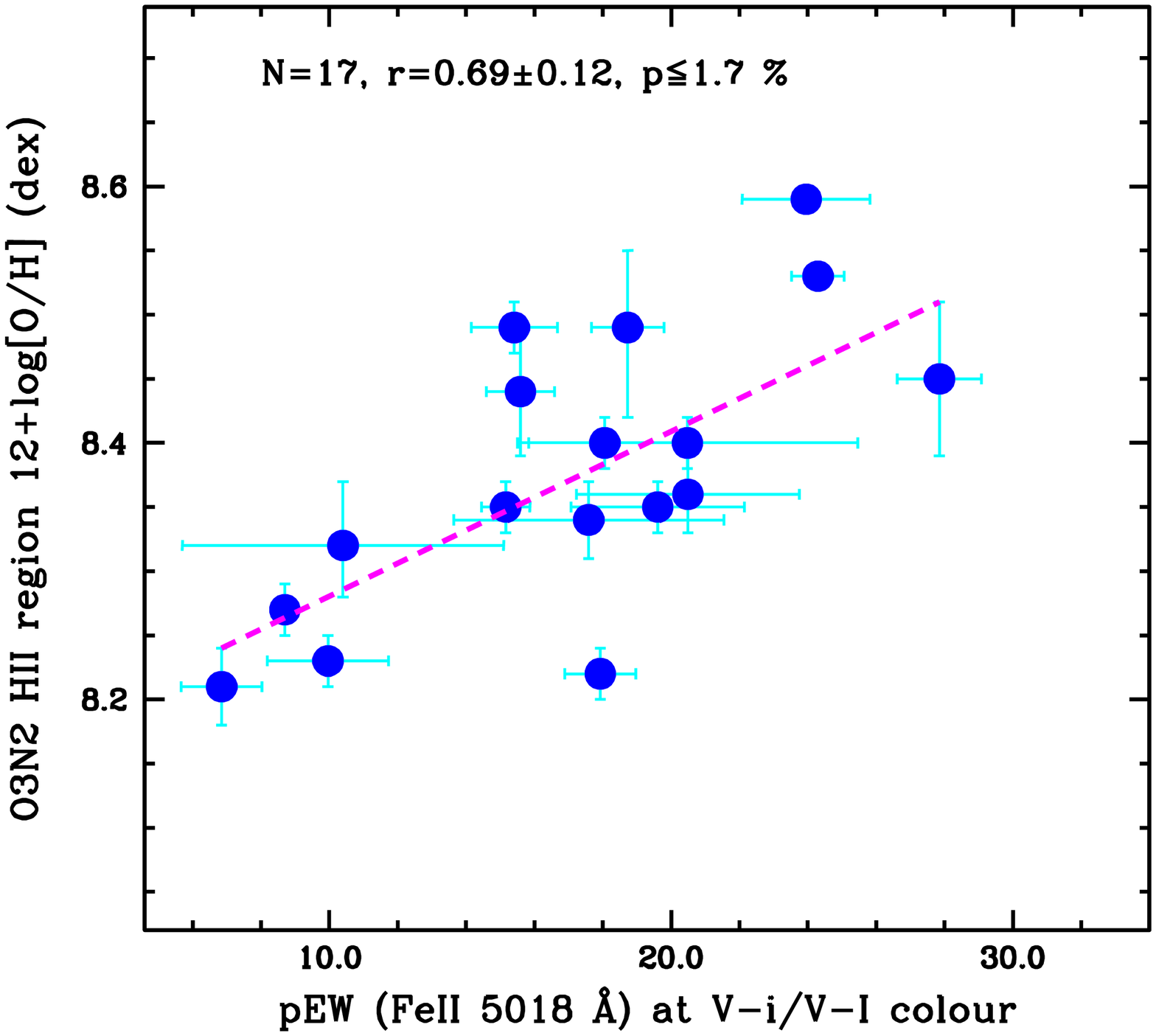}
\caption{SN~II pEWs against O3N2 abundances for the $V-i$/$V-I$ colour sample. }
\label{goldcol}
\end{figure}

\indent Now, in place of time, we investigate whether a stronger correlation exists
if pEWs are measured at a colour epoch.  As already seen in the D14 spectral models: differences
in pre-SN properties can significantly affect the strength of spectral lines (see Figs.~\ref{ewtimebin},~\ref{n2models}). Together with the 
metal abundance within the ejecta, the other main contributor to line appearance and subsequent strength 
is the temperature/ionisation of the line formation region. 
Models corresponding to different progenitor radii produce SNe~II
with distinct temperature evolution. Therefore, examining observational pEWs at a consistent colour (i.e. temperature)
may be of interest.
\begin{figure}
\includegraphics[width=9cm]{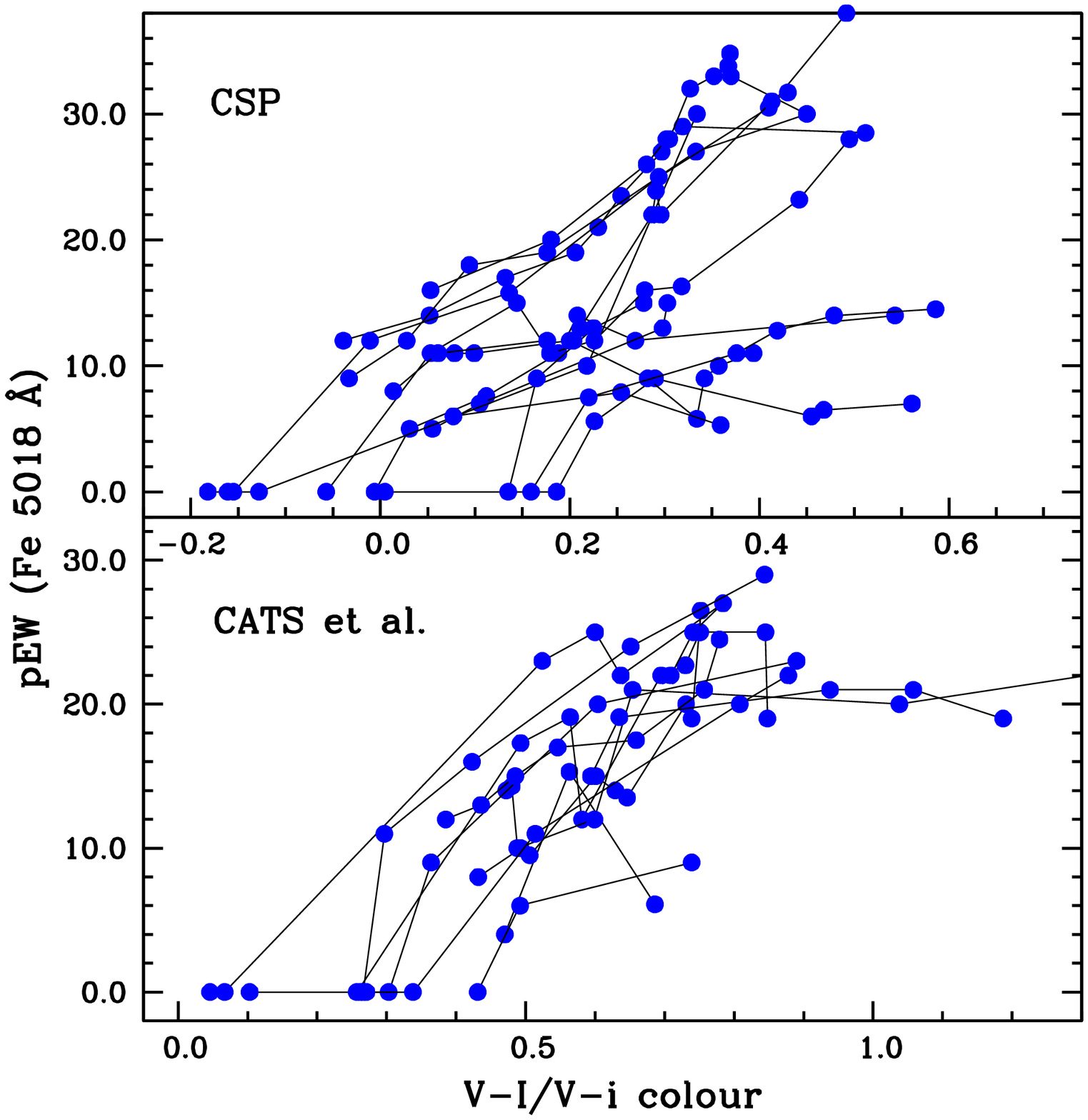}
\caption{Evolution of \feii\ 5018\,\AA\ pEWs with SN colour. In
the top panel the CSP sample is shown using the $V-i$ colour, while in the bottom
panel the CATS et al. sample using $V-I$ is presented. (Errors on colours and
pEWs are not presented to enable better visualisation of trends.)}
\label{colew}
\end{figure}
Fig.~\ref{colew} presents the evolution of pEWs with SN
colour (using the sample defined below). This confirms the behaviour discussed
above: pEWs increase with increasing SN colour.\\
\indent A major issue with any colour analysis is correction for extinction in the line-of-sight 
within host galaxies. However, as discussed in detail in \cite{far14a}, without detailed modelling
and early-time data (see e.g. \citealt{des08}), there is no current satisfactory method for
accurately correcting SNe~II for host galaxy reddening. This is particularly pertinent for the current study
where our goal is to obtain differences in intrinsic colours (i.e. assuming SNe~II have similar colours during the
plateau would simply lead us to use a time epoch). Another complication is that 
the CSP and previous CATS et al. samples do not have the same filter observations.\\
\indent To continue the investigation we proceed in the following manner.
To begin, all SNe~II having host galaxy $A_{\rm V}$ values published in A14 higher than 0.1 magnitude, or where
no $A_{\rm V}$ estimate was possible because either the pEW of narrow ISM sodium lines was higher than 1\,\AA, or
where the upper limit of this quantity was higher than 1\,\AA, are removed from the sample. We then assume that 
the rest of the sample is effectively free of significant host galaxy reddening.  
For all SNe~II within the CSP sample we create $V-i$ colour curves, 
while for the CATS et al. samples $V-I$ colour curves (both corrected for MW extinction) are produced\footnote{Note, CATS et al. photometry 
was published in \citealt{gal16}, while CSP optical photometry will be
published in Anderson et al. (in preparation), and a full colour analysis of the latter
concentrating on host galaxy extinction will be provided in de Jaeger et al. (in preparation).}. 
Low-order
polynomials are fit to these colour curves and we then interpolate to measure the colour at 50\,d.
For each sample (CSP/CATS et al.) a mean 50 day $V-i$/$V-I$ colour is calculated. For each 
individual SN we proceed to measure a pEW for \feii\ 5018\,\AA\ at the corresponding mean colour epoch. This method negates any need to convert $i$-band magnitudes into $I$, or vice versa, 
and is valid if it is assumed that the CSP and CATS et al. samples are drawn from the same underlying SN~II distribution.\\
\indent In Fig.~\ref{goldcol} we plot \feii\ 5018\,\AA\ pEWs at these colour epochs against host \hii-region abundance
on the O3N2 scale. Testing for correlation we find, for N = 17, $r$ = 0.69$\pm$0.12 and $p$ $\leq$1.7 \%, while for the 
same SNe~II but with pEWs measured at 50\,d we find $r$ = 0.61$\pm$0.16 and $p$ $\leq$7.0 \%.
This suggests that in using a colour epoch in place of time, one further removes SN systematics and strengthens 
the case for SNe~II to be used as metallicity indicators.

\begin{table*}
\centering
\caption{Statistics of correlations between \feii\ 5018\,\AA\ pEWs and SN~II light-curve and spectral parameters. 
In the first column the SN~II parameter is listed. (M$_{max}$ is the maximum $V$-band absolute magnitude; M$_{end}$ the magnitude at the 
end of the `plateau', M$_{tail}$ the magnitude at the start of the radioactive tail; s$_1$ the initial
decline from maximum; s$_2$ the decline rate during the `plateau'; s$_3$ the decline rate of the radioactive
tail; $^{56}$Ni the synthesised nickel mass, with $^{56}$Ni* including upper limit calculations; 
Pd the duration from the inflection point of s$_1$ and s$_2$ to the end of the `plateau'; OPTd the duration from explosion 
to the end of the `plateau'; a/e the ratio of the pEW of absorption to emission of \ha; and \ha$_{vel}$ the 
FWHM velocity of \ha.)
This is followed by the number of SNe~II in column 2. 
The Pearson's mean $r$ value is then given together
with the standard deviation. Finally in column 4 the chance probability of finding
a correlation is listed.}
\begin{tabular}[t]{cccc}
\hline
LC/spec parameter & N SNe~II & $r$ & $p$ \\
\hline
M$_{max}$ & 77 & 0.64$\pm$0.09 & $\leq$ 2.2$\times$10$^{-4}$ \%\ \\
M$_{end}$ & 79 & 0.47$\pm$0.11 & $\leq$ 0.11 \%\ \\
M$_{tail}$& 29 & 0.48$\pm$0.14 & $\leq$ 7.1 \%\ \\
s$_1$     & 20 & --0.48$\pm$0.15 & $\leq$ 16 \%\ \\
s$_2$     & 77 & --0.41$\pm$0.13 & $\leq$ 1.4 \%\ \\
s$_3$     & 24 & --0.50$\pm$0.24 & $\leq$ 22 \%\ \\
$^{56}$Ni & 13 & --0.46$\pm$0.23 & $\leq$ 45 \%\ \\
$^{56}$Ni*& 36 & --0.54$\pm$0.10 & $\leq$ 0.72 \%\ \\
Pd        & 17 & 0.26$\pm$0.31 & $\leq$ 100 \%\ \\
OPTd      & 24 & 0.12$\pm$0.14 & $\leq$ 100 \%\ \\
a/e		  & 43 & 0.60$\pm$0.10 & $\leq$ 0.064 \%\ \\
\ha$_{vel}$    & 24 & --0.61$\pm$0.09 & $\leq$ 0.92 \%\ \\
\hline
\hline	
\end{tabular}
\label{ewlcspec}
\end{table*}

\subsection{Correlations between pEWs and other SN~II parameters}
Here we correlate pEWs at 50\,d
with the SN~II light-curve and spectral parameters presented by
A14 and \cite{gut14}. The statistical significance of correlations between
pEWs and these parameters are listed in Table~\ref{ewlcspec}.
A full discussion of these correlations, together with figures and discussion of 
their implications 
for our understanding of SNe~II explosions and progenitors will be left for
a future publication (Gut\'ierrez et al. in preparation). However, there are some interesting trends seen in Table~\ref{ewlcspec}
and we briefly discuss those now.\\
\indent The absolute magnitude at maximum light, M$_{max}$, of SN~II is found to strongly 
correlate with the \feii\ 5018\,\AA\ pEW at 50\,d. Indeed, while the pEW also shows correlation with M$_{end}$ and M$_{tail}$
(all in the sense that brighter SNe~II have lower pEWs at a consistent time epoch), the strength of the 
correlation is less. This is in agreement with A14, where M$_{max}$ was shown to be a more important 
parameter in understanding the diversity of SNe~II than the end of `plateau' magnitude M$_{end}$.
The decline rates, s$_1$, s$_2$ and s$_3$, all show some degree of correlation with the \feii\ 5018\,\AA\ pEW
in the direction that slower decliners have higher pEWs.  
Interestingly, when we include upper limits, $^{56}$Ni masses show a strong correlation with pEWs: SNe~II which produce more nickel
have lower pEWs. The time duration OPTd, i.e. the time between explosion and the end of the `plateau' together with Pd ($t_{\rm tran}$
to end of `plateau') show
zero correlation with pEWs.
Finally, the spectral parameters a/e (the ratio of the pEW of absorption to emission of \ha) and the FWHM velocity of \ha\ (see
\citealt{gut14} for more details) both show a strong correlation with pEWs. SNe~II with large pEWs at 50\,d have larger a/e values
and smaller velocities.

\subsection{The influence of metallicity on SN~II diversity}
\begin{figure}
\includegraphics[width=9cm]{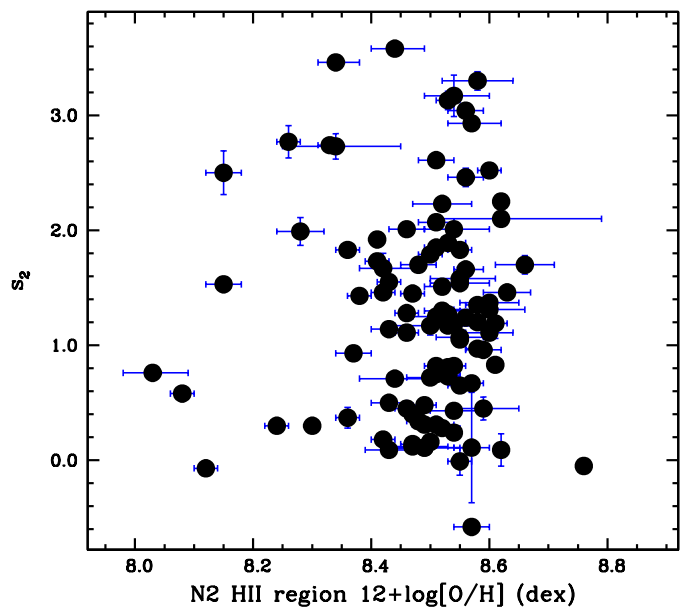}
\caption{Host \hii-region oxygen abundance plotted against SN~II `plateau' decline, s$_{2}$.}
\label{figZs2}
\end{figure}

\begin{figure}
\includegraphics[width=9cm]{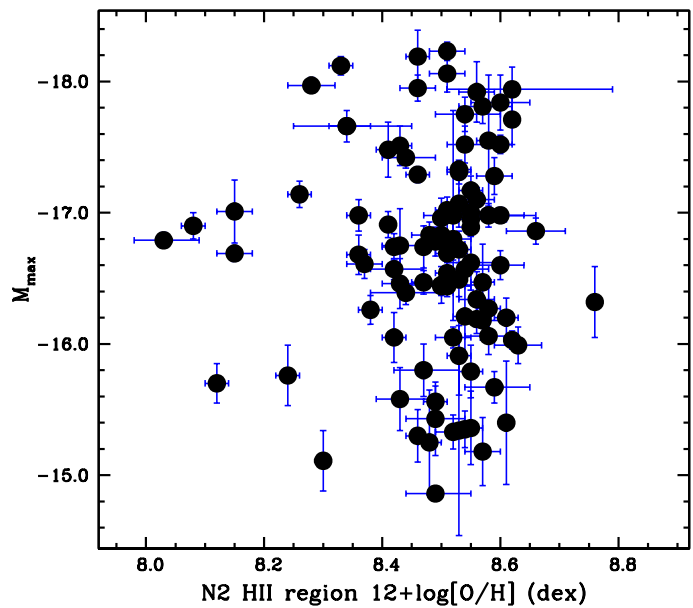}
\caption{Host \hii-region oxygen abundance plotted against SN absolute $V$-band
maximum, M$_{max}$.}
\label{figZmax}
\end{figure}

\begin{figure}
\includegraphics[width=9cm]{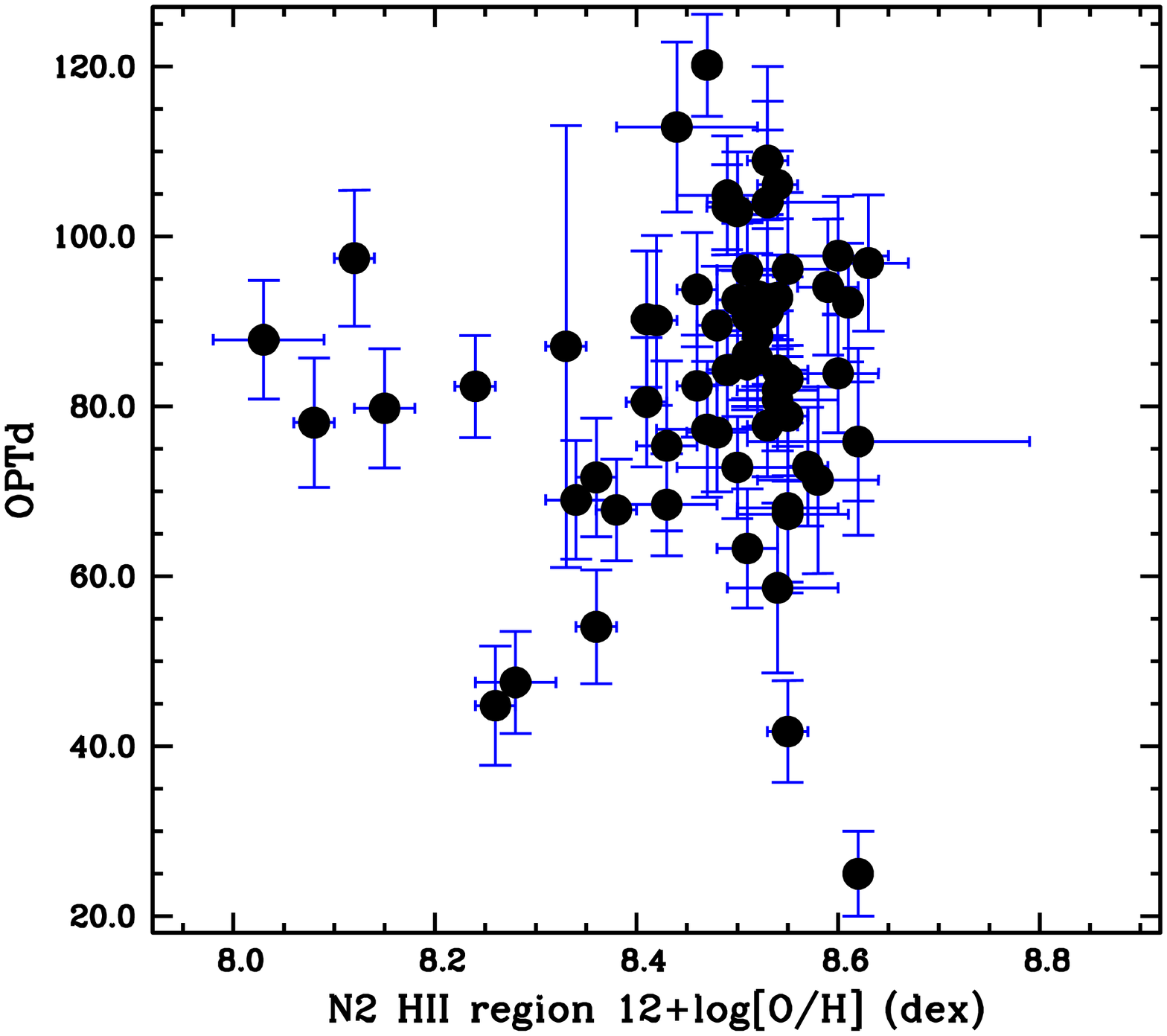}
\caption{Host \hii-region oxygen abundance plotted against SN OPTd: time duration
from explosion to end of `plateau'.}
\label{figZoptd}
\end{figure}

We test for correlation between SN~II host \hii-region abundance and light-curve
and spectral parameters.
No evidence for correlation is found between abundance and any
SN~II parameter, except that with metal line pEWs. In Figs.~\ref{figZs2},~\ref{figZmax},~\ref{figZoptd} we show
host \hii-region abundance plotted against M$_{max}$, s$_2$, and OPTd respectively.
One can see that no trends appear. In Table~\ref{Zlcspec} the results of statistical tests
for trends between these parameters and host \hii-region abundance, together with all other SN~II
light-curve and spectral properties presented in A14 and \cite{gut14} are shown. In addition, we repeat the statistics for the 
correlation between SN~II \feii\ 5018\,\AA\ pEWs and host \hii\ region abundance already presented above.
It is clear that the only (thus far measured/presented) SN~II parameter which shows any correlation
with environment metallicity is the \feii\ 5018\,\AA\ pEW.\\
\indent 
These results suggest that the diversity of SN II properties in 
the current sample does not stem from variations in metallicity -- either the range in 
metallicity is too small to matter (which is likely) or some other 
stellar parameter (e.g. main sequence mass) is more influential.

%These results suggest that at least in the current sample progenitor metallicity does not have a significant effect 
%on the evolution of progenitor stars in terms of significantly altering their pre-SN 
%properties, which are then reflected in the diversity of SNe~II observed.

\begin{table*}
\centering
\caption{Statistics of correlation tests between host \hii-region abundance, 
and SN~II light-curve and spectral parameters. 
In the first column the SN~II parameter is listed (described in the caption of Table~\ref{ewlcspec}), 
followed by the number of SNe~II in column 2.
The Pearson's mean $r$ value is then given together
with the standard deviation. Finally in column 4 the chance probability of finding
a correlation is listed}
\begin{tabular}[t]{cccc}
\hline
LC/spec parameter & N SNe~II & $r$ & $p$\\
\hline
M$_{max}$ & 102& 0.04$\pm$0.08 & $\leq$ 100 \%\ \\
M$_{end}$ & 104& 0.09$\pm$0.08 & $\leq$ 92 \%\ \\
M$_{tail}$& 37 & 0.003$\pm$0.16 & $\leq$ 100 \%\ \\
s$_1$     & 24 & --0.17$\pm$0.30 & $\leq$ 100 \%\ \\
s$_2$     & 102& --0.02$\pm$0.11 & $\leq$ 100 \%\ \\
s$_3$     & 28 & --0.27$\pm$0.28 & $\leq$ 100 \%\ \\
$^{56}$Ni*& 43 & 0.12$\pm$0.16 & $\leq$ 100 \%\ \\
Pd        & 18 & 0.26$\pm$0.31 & $\leq$ 100 \%\ \\
OPTd      & 66 & 0.12$\pm$0.14 & $\leq$ 100 \%\ \\
a/e		  & 43 & 0.23$\pm$0.13 & $\leq$ 52 \%\ \\
\ha$_{vel}$& 43 & 0.009$\pm$0.12 & $\leq$ 100 \%\ \\
\textbf{\feii\ 5018\,\AA\ pEW (50)} & \textbf{82} & \textbf{0.34$\pm$0.09} & \textbf{$\leq$ 2.4 \%} \\
\hline
\hline	
\end{tabular}
\label{Zlcspec}
\end{table*}

%__________________________________________________________________

\section{Discussion}
\cite{des14} presented SN~II model spectra produced by progenitors of distinct metallicity, and showed
how the strength of metal lines increases with increasing progenitor metallicity. 
Those models therefore made a prediction that SNe~II with higher metal line pEWs
will be found in environments of higher metallicity.
In this publication
we have concentrated on the strength of the \feii\ 5018\,\AA\ line as measured in observed SN~II spectra,
and shown that indeed the pEW of this line shows a statistically significant trend with 
host \hii-region abundance, with the latter derived from the ratio of \hii-region emission lines.
We now discuss this result in more detail, and outline the next steps 
to use SNe~II as independent environment metallicity probes. The implications
of the above results on the progenitors and pre-SN evolution of SNe~II are also further explored.

\subsection{The use of photospheric phase SN~II spectra as independent metallicity indicators}
The aim of this work is to confront model predictions with observations to probe the 
accuracy of SNe~II as metallicity indicators. The correlation observed between
SN \feii\ 5018\,\AA\ pEW and host \hii-region abundance suggests that SNe~II may be at least
as accurate at indicating environment metallicity as the popularly used N2 and O3N2 diagnostics.
This is inferred from Figs.~\ref{ew50_all},~\ref{O3N2all},~\ref{goldII}, and ~\ref{goldcol}. The dispersion within 
these correlations is similar to (or better than) the internal dispersion of the emission line diagnostics.
This probably implies that a) the majority of the dispersion on the pEW--abundance plots arises from 
the diagnostics and not the SN measurements, and b) the RMS values of M13 underestimate the 
true precision of N2 and O3N2. This finding motivates work to use SNe~II 
as metallicity indicators independent of these emission line diagnostics.\\
\indent Currently there is a lack of SNe~II within this sample at both low (sub--LMC) and high
(super-solar) metallicities (although we
caution that the latter may be due to the saturation of the M13 diagnostics
at above $\sim$solar metallicity). Finding SNe~II within these environments and adding these to the sample
would allow metallicity differences to dominate changes in pEW, and lend further
support to the use of SN~II as metallicity probes. A first step to remove the calibration from strong
emission line diagnostics could be to obtain deeper spectra of the host \hii\ regions to 
enable detections of emission lines which provide electron temperature estimates, i.e.
the direct method. In this case SNe~II would still be tied to the same scale 
as those diagnostics used throughout the Universe. One may look at models to calibrate
the SN observations. However, to do this accurately would probably require individual fitting
of each SN to a large grid of spectral models. Currently, what we can state 
with confidence is
that an observed SN~II at $\sim$50\,d days post explosion with a pEW $<$10\,\AA\ suggests an environment
metallicity of less than or equal to LMC metallicities ($\lesssim$0.2\zsun). At the other
extreme, if one finds a SN~II with an \feii\ 5018\,\AA\ pEW $\gtrsim$ 30\,\AA, this implies
the SN exploded within a $\sim$solar abundance (12+log[O/H] = 8.69 dex, \citealt{asp09}) region or higher.

\subsection{Progenitor metallicity as a minor player in producing SN~II light-curve and spectral diversity?}
Metallicity is thought to be a key ingredient in stellar evolution, driving the extent of mass-loss
through metallicity dependent stellar winds. However, a major finding presented here is of
zero evidence for correlation between SN parameters and environment abundance (except in the case of 
the pEW of metal lines affected by the nascent composition of the SN ejecta). 
This suggests metallicity is in fact playing a negligible role in 
producing the diversity observed \textit{in the current sample}. However, we suspect the range in metallicity is too narrow to drive SN II diversity. There are many other
parameters which likely change within our sample such as progenitor mass, mass-loss rates, degree of
binary interaction etc. It may be that when we eventually probe a larger range of metallicity we 
start to see the effects of progenitor metallicity on SN diversity. It is noted however that even 
if our sample lacks SNe~II in the extremes of the metallicity distribution, the rate of these events is unlikely
to be significant compared to the current sample. In conclusion, 
over the range of progenitor metallicities we have in our current sample, chemical abundance 
\textit{is not} a driving force in producing SN~II light-curve and spectral diversity.\\

\subsection{The lack of SNe~II in low-metallicity environments}
The lack of SNe~II in low metallicity environments (e.g. those found in the Small Magellanic Cloud or lower)
was already noted in \cite{sto13} and D14. In the current analysis we now have additional constraints through our environment
abundances. In a sample of 115 (52) SNe~II on the N2 (O3N2) scale, the lowest metallicity environment is that of SN~2003cx, at an
oxygen abundance of 8.03 (8.06) dex on the N2 (O3N2) scale. Using the solar abundance of 8.69 dex \citep{asp09} 
this translates to 0.22\zsun, i.e.
$\sim$abundances found in the LMC.
As noted above and discussed in \cite{lop12}, the N2 and O3N2 scales give systematically lower abundances than those from other emission line 
methods. If this is the case then the lack of low-metallicity environment SNe~II would become even more apparent.
There also appears to be a lack of SNe~II in super-solar metallicity environments. However, if we accept that 
the N2 and O3N2 scales give
systematically low abundances, then this may become less significant.\\
\indent Other SN~II environment metallicities have been published. For sake
of comparison we calculate our distributions now on the \cite{pet04} scale (see Table~\ref{hiilist}).
On this scale a mean N2 abundance $=$ 8.62$\pm$0.21 dex and a mean O3N2 abundance = 8.55$\pm$0.21 dex is found.
\cite{and10} published a sample of 46 SNe~II oxygen abundances and derived a very similar distribution
as that found here, which is unsurprising given that about half of those values are included in our sample.
\cite{kun13_2} analysed very nearby SNe~II environments and found a mean 12+log[O/H] value of 8.58 dex for 15 SNe~II,
again very similar to our sample. One common parameter between these samples is that the SNe~II were generally found
via galaxy targeted searches, and therefore they may be biased towards more massive, higher metallicity galaxies.
Indeed, the Palomar Transient Factory (PTF, \citealt{rau09}) published a distribution of CC SN host galaxy absolute
magnitudes which probed a greater number of dwarf galaxies than found in, e.g. our sample (see further discussion in 
the appendix of A14). However, \cite{sto13} obtained emission line spectra of the host \hii\ regions of a representative
sample of PTF SN~II hosts, and concluded that the abundance distribution was indistinguishable from 
that found in targeted samples (e.g. \citealt{and10}). In conclusion, there appears to be a true lack of (published)
SNe~II found with SMC or lower metallicities. Finding and studying such SNe~II will not only serve our analysis 
to further calibrate SNe~II as metallicity probes, but will also allow the study of how massive stars explode at low
metallicity.

\subsection{Future directions}
In this work we have focussed on the dependence of the pEW of a single spectral line \feii\ 5018\,\AA\ on
environment metallicity. In future work we will analyse the full distribution of SN photospheric phase
metal line strengths to determine the most direct indicator of progenitor metallicity.
In addition, we are currently limited to a relatively small
range in metallicity. Observing additional SNe~II outside of this range
will aid in removing other
systematic uncertainties in the correlations we have presented.\\
\indent While there is still much to understand, the ultimate aim of this work would be to independently 
map the metallicity distribution of galaxies throughout the Universe using SNe~II. To aid in this goal one may think of
observing SNe~II in both a range of environments and out to higher redshifts. The advantage of SNe~II
over traditional metallicity indicators is that a) they probe the specific location where they explode 
within their hosts, b) in principle one can obtain abundances of distinct elements and not simply
oxygen, and c) SNe~II are intrinsically bright thus the currently proposed
metallicity diagnostic can be efficient -- in terms of 
telescope observing time -- for probing distant galaxies. 
In terms of constraining a range of relative metal
abundances it is important to calibrate all metal lines within SNe~II
spectra, and understand the systematics in their measurements and metallicity predictions. 
SNe~II explosions accurately trace the star formation within galaxies (e.g. \citealt{bot12}). Hence, 
mapping metallicity with SNe~II discovered by un-targeted searches will accurately trace the 
chemical abundance of star-forming regions within galaxies.

%__________________________________________________________________

\section{Conclusions}
Following the study of D14, we present observations of a large sample
of SN~II host \hii-region spectroscopy from which gas-phase oxygen abundances are inferred. These
are compared to pEW measurements of \feii\ 5018\,\AA\ and a statistically significant trend is observed, in that
SNe~II with higher pEWs explode in higher metallicity environments. 
This paves the way
for the use of SNe~II as independent metallicity indicators throughout the Universe.
While we observe significant dispersion in this trend, this is expected because a) the SNe~II included show many
different properties, and b) the abundance diagnostic used for comparison itself shows significant dispersion
in its correlation with electron temperature. Indeed, the significance of correlation
is increased if we only consider: a) \hii-region measurements close to explosion sites; b)
SNe~II which have light-curve morphologies similar to `normal' SNe~IIP; and c) if we use a colour epoch for pEW
measurements in place of time.\\
\indent We also search for trends of progenitor (inferred from environment) metallicity with various
SN~II light-curve and spectral parameters. However, no such trends are observed.
We therefore conclude, that at least within the current sample, 
progenitor metallicity plays a negligible role in producing the observed diversity of
SNe~II.\\

\begin{acknowledgements}
The annonymous referee is thanked for their useful comments which helped
clarify some important points in the paper.
Support for C.~G., M.~H., L.~G. and H.~K is provided by the Ministry of Economy, Development, and
Tourism's Millennium Science Initiative through grant IC120009, awarded to
The Millennium Institute of Astrophysics, MAS.
L~.G. and H~.K. acknowledge support by CONICYT through FONDECYT grants 3140566 and 3140563 respectively.
L.~D. acknowledges financial support from ``Agence Nationale de la Recherche''
grant ANR-2011-Blanc-BS56-0007.
The work of the CSP has been supported by the National Science Foundation under grants AST0306969, AST0607438, and AST1008343.
M. Stritzinger gratefully acknowledges the generous support provided by the Danish Agency for Science and Technology and Innovation  
realized through a Sapere Aude Level 2 grant.
Based on observations made
with ESO telescopes at the La Silla Paranal Observatory under pro-
gramme: 094.D-0283(A).
A special thanks goes to the ANTU support staff at Paranal observatory
for obtaining the observations used in this publication. In particular
we acknowledge the telescope and instrument operators: Israel Blanchard, Claudia Cid, Alex Correa,
Lorena Faundez, Patricia Guajardo, Diego Parraguez, Andres Parraguez, Marcelo Lopez, Julio Navarrete, Leonel Rivas, Rodrigo Romero, 
and Sergio Vera.
This research
has made use of the NASA/IPAC Extragalactic Database (NED) 
which is operated by the Jet 
Propulsion Laboratory, California
Institute of Technology, under contract with the National Aeronautics.
\end{acknowledgements}

%-------------------------------------------------------------------

\bibliographystyle{aa}

\bibliography{Reference}

\appendix

\section{SN and host galaxy data}
The details of all SNe~II and their host galaxies included in the current analysis are presented in Table A.1. 
These SNe~II are those events analysed in A14 and \cite{gut14}, together
with several other SNe~II from the CSP et al. surveys. In addition, we obtained host \hii-region spectroscopy
of a few SNe~IIb and a couple of SNe~IIn, the details of which are also listed in Table A.1.

\clearpage

\begin{table*}
\centering
\caption{SN and host galaxy data}
\begin{tabular}[t]{cccccccccc}
\hline
SN & Host galaxy & $V_{\rm r}$(km s$^{-1}$) & Host $M_{\rm B}$ & $s_{\rm 2}$(mag 100d$^{-1}$) & $M_{\rm max}$(mag) & OPTd(d) & a/e & \ha$_{\rm vel}$(km s$^{-1}$) \\
\hline	
\hline
1986L	&NGC 1559   &1305   &--21.3 &1.28 &--18.19 &93.7    &0.21 &6354 \\
1990E	&NGC 1035   &1241   &--19.2 &$\cdots$  &$\cdots$     &$\cdots$     &$\cdots$  &$\cdots$  \\			
1990K	&NGC 150    &1588   &--20.2  &$\cdots$  &$\cdots$     &$\cdots$     &$\cdots$  &$\cdots$	\\	
1991al	&NGC 4411B  &4575   &--18.8 &1.55 &--17.51 &$\cdots$&0.28 &7771 \\
1992af	&ESO 340-G038&5541  &--19.7 &0.37 &--16.98 &54.03	&$\cdots$&$\cdots$\\
1992am	&MCG --01-04-039&14397&--21.4&1.17&--17.33 &$\cdots$&$\cdots$&$\cdots$\\
1992ad	&NGC 4411B     &1270   &--18.6    &$\cdots$ &$\cdots$     &$\cdots$     &$\cdots$       &$\cdots$\\
1992ba	&NGC 2082   &1185   &--18.0 &0.73 &--15.34 &103.97	&0.68	&4439\\
1993A	&anon       &8790   &$\cdots$    &0.72 &--16.44 &$\cdots$&$\cdots$&$\cdots$\\
1993K	&NGC 2223   &2724   &--20.9 &2.46 &--17.92 &$\cdots$&$\cdots$&$\cdots$\\
1993S	&2MASX J22522390&9903&--20.6&2.52&--17.52&$\cdots$&$\cdots$&$\cdots$\\
1999br	&NGC 4900   &960    &--19.4 &0.14 &--13.77 &$\cdots$&0.61  &3566\\
1999ca	&NGC 3120   &2793   &--20.4 &1.73 &--17.48 &80.48	&$\cdots$&$\cdots$\\
1999cr	&ESO 576-G034&6069  &--20.4 &0.58 &--16.90 &78.06	&0.19	&5728\\
1999eg	&IC 1861    &6708   &--20.9 &1.70 &--16.86 &$\cdots$&$\cdots$&$\cdots$\\
1999em	&NGC 1637   &717    &--19.1 &0.31 &--16.76 &96.04	&0.57	&5915\\
2002ew	&NEAT J205430.50&8975&$\cdots$   &3.58 &--17.42 &$\cdots$	&$\cdots$	    &$\cdots$\\
2002fa	&NEAT J205221.51&17988&$\cdots$  &1.58 &--16.95 &67.29	&$\cdots$	    &$\cdots$\\
2002gd	&NGC 7537   &2676   &--19.8 &0.11 &--15.43 &$\cdots$&0.19	&4023\\
2002gw	&NGC 922    &3084   &--20.8 &0.30 &--15.76	&82.33	&0.46	&6217\\
2002hj	&NPM1G +04.0097&7080&$\cdots$    &1.92 &--16.91	&90.24	&0.38	&6857\\
2002hx	&PGC 023727 &9293   &$\cdots$    &1.54 &--17.00	&68.03	&$\cdots$	    &$\cdots$\\
2002ig	&anon       &23100  &$\cdots$    &2.73 &--17.66	&$\cdots$	&	$\cdots$    &$\cdots$\\
2003B 	&NGC 1097   &1272   &--21.4 &0.65 &--15.36	&83.19	&0.4	&4251\\
2003E	&MCG--4-12-004 &4470&--19.7 &-0.07&--15.70	&97.42	&0.40	&5028\\
2003T	&UGC 4864   &8373   &--20.8 &0.82 &--16.54	&90.59	&0.55	&7360\\
2003bl	&NGC 5374   &4377   &--20.6 &0.24 &--15.35	&92.81	&0.47	&6596\\
2003bn	&2MASX J10023529&3828&--17.7&0.28 &--16.80	&92.97	&0.6	&6121\\
2003ci	&UGC 6212   &9111   &--21.8 &1.79 &--16.83	&92.53	&$\cdots$	    &$\cdots$\\
2003cn	&IC 849     &5433   &--20.4 &1.43 &--16.26	&67.80	&0.22	&5074\\
2003cx	&NEAT J135706.53&11100&$\cdots$  &0.76 &--16.79	&87.82	&0.29	&7314\\
2003dq	&MAPS-NGP O43207&13800&$\cdots$  &2.50 &--16.69	&$\cdots$	&$\cdots$	    & $\cdots$   \\
2003ef	&UGC 7820   &5094   &--20.1 &0.81 &--16.72	&90.93	&	$\cdots$    &   $\cdots$ \\
2003eg	&NGC 4727   &4388   &--22.3 &2.93 &--17.81	&$\cdots$	&$\cdots$	    & $\cdots$   \\
2003ej	&UGC 7820   &5094   &--20.1 &3.46 &--17.66	&68.97	&	$\cdots$    &$\cdots$\\
2003fb	&UGC 11522  &5262   &--20.9 &0.48 &--15.56	&84.27	&	$\cdots$    &$\cdots$\\
2003gd	&M74        &657    &--20.6 &$\cdots$	&$\cdots$	&$\cdots$	&	$\cdots$    &$\cdots$\\
2003hd	& MCG-- 04-05-010&11850&--21.7&1.11	&--17.29	&82.39	&0.76	&4800\\
2003hk	&NGC 1085   &6795   &--21.3 &1.85	&--17.02	&86.00	&$\cdots$	    &$\cdots$\\
2003hl	&NGC 772    &2475   &--22.4 &0.74	&--15.91	&108.92	&	$\cdots$    &$\cdots$\\
2003hn	&NGC 1448   &1170   &--21.1 &1.46	&--16.74	&90.10	&0.29	&7268\\
2003ho	& ESO 235-G58&4314  &--19.8 &$\cdots$	&$\cdots$	&$\cdots$	&$\cdots$	    & $\cdots$   \\
2003ib	& MCG-- 04-48-15 &7446 &--20.8&1.66	&--17.10	&$\cdots$	&$\cdots$	    &  $\cdots$  \\
2003ip	&UGC 327    &5403   &--19.4 &2.01	&--17.75	&80.74	&	$\cdots$    &$\cdots$\\
2003iq	&NGC 772    &2475   &--22.4 &0.75	&--16.69	&84.91	&	$\cdots$     &$\cdots$ \\
2004dy	&IC 5090    &9352   &--20.9 &0.09	&--16.03	&24.96	&	$\cdots$     &$\cdots$ \\
2004ej	&NGC 3095   &2723   &--20.9 &1.07	&--16.62	&96.14	&	 $\cdots$   &$\cdots$\\
2004er	&MCG-- 01-7-24 &4411&--20.2 &0.40	&--16.74	&120.15	&0.56	&7680\\
2004fb	&ESO 340-G7 &6100   &--20.9 &1.24	&--16.19	&$\cdots$	&$\cdots$	    &$\cdots$\\
2004fc	&NGC 701    &1831   &--19.5 &0.82	&--16.21	&106.06	&0.37	&5440\\
2004fx	& MCG-- 02-14-3 &2673&$\cdots$&0.09	&--15.58	&68.41	&$\cdots$	    &$\cdots$\\
2005J	&NGC 4012   &4183   &--20.4 &0.96	&--17.28	&94.03	&0.54	&6637\\
2005K	&NGC 2923   &8204   &--19.6&1.67	&--16.57	&$\cdots$	&$\cdots$	    &  $\cdots$  \\
2005Z	&NGC 3363   &5766   &--19.6&1.83	&--17.17	&78.84	&	$\cdots$    &$\cdots$\\
\hline                      
\hline                      
\end{tabular}               
\end{table*}

\begin{table*}
\centering
\setcounter{table}{0}
\begin{tabular}[t]{cccccccccc}
\hline
SN & Host galaxy & $V_{\rm r}$(km s$^{-1}$) & Host $M_{\rm B}$ & $s_{\rm 2}$(mag 100d$^{-1}$) & $M_{\rm max}$(mag) & OPTd(d) & a/e & \ha$_{\rm vel}$(km s$^{-1}$) \\
\hline	
\hline
2005af	&NGC 4945   &563    &--20.5 &$\cdots$	&$\cdots$	&104.01	&	$\cdots$    &$\cdots$\\
2005an	&ESO 506-G11&3206   &--18.6&1.89	&--17.07	&77.71	&0.17	&8548\\
2005dk	&IC 4882    &4708   &--19.8 &1.18	&--17.52	&84.22	&0.3	&7008\\
2005dn	&NGC 6861   &2829   &--21.0 &1.53	&--17.01	&79.76	&$\cdots$	    &$\cdots$\\
2005dt	&MCG --03-59-6&7695 &--20.9 &0.71	&--16.39	&112.86	&	$\cdots$    &$\cdots$\\
2005dw	&MCG --05-52-49&5269&--21.1 &1.27	&--16.49	&92.59	&	$\cdots$     &$\cdots$ \\
2005dx	&MCG --03-11-9&8012 &--20.8&1.30	&--16.05	&85.59	&	 $\cdots$    &$\cdots$ \\
2005dz	&UGC 12717  &5696   &--19.9     &0.43	&--16.57	&81.86	&0.66	&5952\\
2005es	&MCG +01-59-79&11287&--21.1     &1.31	&--16.98	&$\cdots$	&	$\cdots$    &  $\cdots$  \\
2005gk	&2MASX J03081572&8773&$\cdots$&1.25	&--16.44	&$\cdots$	&$\cdots$	    &$\cdots$\\
2005gz	&MCG -01-53-22&8518&--21.3&$\cdots$		&$\cdots$		&$\cdots$       &$\cdots$       &$\cdots$\\
2005kh	&NGC 3094&2220&--19.7&$\cdots$		&$\cdots$		&$\cdots$       &$\cdots$       &$\cdots$\\
2005me	&ESO 244-31 &6726   &--21.4     &1.70	&--16.83	&76.91	&	$\cdots$     &$\cdots$ \\
2006Y	&anon       &10074  &$\cdots$        &1.99	&--17.97	&47.49	&0.01	&7588\\
2006ai	&ESO 005-G009&4571&--19.2       &2.07	&--18.06	&63.26	&0.08	&7291\\
2006bc	&NGC 2397   &1363   &--20.9     &-0.58	&--15.18	&$\cdots$	&	$\cdots$    &  $\cdots$  \\
2006be	&IC 4582    &2145   &--18.7     &0.67	&--16.47	&72.89	&0.34	&6308\\
2006bl	&MCG +02-40-9&9708  &--20.9     &2.61	&--18.23	&$\cdots$	&	$\cdots$    & $\cdots$   \\
2006it	&NGC 6956   &4650   &--21.2     &1.19	&--16.20	&$\cdots$	&$\cdots$	    &$\cdots$\\
2006iw	&2MASX J23211915&9226&--18.3    &1.05	&--16.89	&$\cdots$	&0.46	&6162\\
2006ms	&NGC 6935   &4543   &--21.3     &0.11	&--16.18	&$\cdots$	&$\cdots$	    & $\cdots$   \\
2006qq*	&	ESO 553-G36&8688&--20.7&$\cdots$		&$\cdots$		&$\cdots$       &$\cdots$       &$\cdots$\\
2006qr	&MCG --02-22-023    &4350&--20.2     &1.46	&--15.99	&96.85	&0.55	&5440\\
2007W	&NGC 5105   &2902   &--20.9     &0.12	&--15.80	&77.29	&0.52	&4800\\
2007X	&ESO 385-G32&2837   &--20.5     &1.37	&--17.84	&97.71	&0.2	&8091\\
2007Z	&PGC 0016993&5333&--22.9&$\cdots$		&$\cdots$		&$\cdots$       &$\cdots$       &$\cdots$    \\
2007aa	&NGC 4030   &1465   &--21.1     &-0.05	&--16.32	&$\cdots$	&0.7	&5028\\
2007ab	&MCG --01.43-2&7056 &--21.5     &3.30	&--16.98	&71.30	&$\cdots$	    &  $\cdots$  \\
2007am**	&NGC 3367&3039&--21.4&$\cdots$		&$\cdots$		&$\cdots$       &$\cdots$       &$\cdots$\\
2007av	&NGC 3279   &1394   &--20.1     &0.97	&--16.27	&$\cdots$	&	    &\\
2007ay**	&UGC 4310&4359&--18.7&$\cdots$		&$\cdots$		&$\cdots$       &$\cdots$       &$\cdots$\\
2007hm	&SDSS J205755&7540  &$\cdots$&1.45	&--16.47	&$\cdots$	&$\cdots$	    &$\cdots$\\
2007il	&IC 1704    &6454   &--20.7     &0.31	&--16.78	&103.43	&0.38	&7634\\
2007it	&NGC 5530   &1193   &--19.6     &1.35	&--17.55	&$\cdots$	&	 $\cdots$   & $\cdots$   \\
2007oc	&NGC 7418   &1450   &--19.9     &1.83	&--16.68	&71.62	&0.11	&7634\\
2007sq	&MCG --03-23-5&4579&--22.2&1.51	&--15.33	&88.34	&	$\cdots$    &$\cdots$\\
2008F	&MCG --01-8-15&5506 &--20.5     &0.45	&--15.67	&$\cdots$	&$\cdots$	    &$\cdots$\\
2008H	&ESO 499-G05&4292&--21.5&$\cdots$		&$\cdots$		&$\cdots$       &$\cdots$       &$\cdots$    \\
2008M	&ESO 121-26 &2267   &--20.4     &1.14	&--16.75	&75.34	&0.22	&6674\\
2008N	&NGC 4273&2382&--20.6 &		&$\cdots$		&$\cdots$       &$\cdots$       &$\cdots$    \\
2008W	&MCG -03-22-7&5757  &--20.7     &1.11	&--16.60	&83.86	&$\cdots$	    &$\cdots$\\
2008ag	&IC 4729    &4439   &--21.5     &0.16	&--16.96	&102.95	&	$\cdots$    &$\cdots$\\
2008aw	&NGC 4939   &3110   &--22.2     &2.25	&--17.71	&75.83	&0.13	&7817\\
2008bh	&NGC 2642   &4345   &--20.9     &1.20	&--16.06	&$\cdots$	&0.22	&6857\\
2008bk	&NGC 7793   &227    &--18.5     &0.11	&--14.86	&104.83	&0.65	&2925\\
2008bm	&CGCG 071-101&9563  &--19.5     &2.74	&--18.12	&87.04	&$\cdots$ 	    & $\cdots$    \\
2008bp	&NGC 3905   &2723   &--21.6     &3.17	&--14.00	&58.62	&$\cdots$ 	    &$\cdots$ \\
2008br	&IC 2522    &3019   &--20.9     &0.45	&--15.30	&$\cdots$	&0.4	&4571\\
2008bu	&ESO 586-G2 &6630   &--21.6     &2.77	&--17.14	&44.75	&	$\cdots$    & $\cdots$   \\
2008fq	&NGC 6907 &3162&--21.8&$\cdots$		&$\cdots$		&$\cdots$       &$\cdots$       &$\cdots$\\
2008ga	& LCSB L0250N&4639  &$\cdots$        &1.17	&--16.45	&72.79	&	 $\cdots$   &$\cdots$\\
2008gi	&CGCG 415-004&7328  &--20.0     &3.13	&--17.31	&$\cdots$	&	$\cdots$    &$\cdots$\\
2008gq**	&MCG -02-26-39&3628&--19.4&$\cdots$&$\cdots$		&$\cdots$		&$\cdots$       &$\cdots$\\
2008gr	&IC 1579    &6831   &--20.6     &2.01	&--17.95	&$\cdots$	&0.17	&8731\\
2008ho	&NGC 922    &3082   &--20.8     &0.30	&--15.11	&$\cdots$	&$\cdots$	    &$\cdots$\\
2008if	&MCG --01-24-10&3440&--20.4     &2.10	&-17.94	&75.85	&0.08	&8731\\
2008il	&ESO 355-G4 &6276   &--20.7     &0.93	&-16.61	&$\cdots$	&$\cdots$	    &  $\cdots$  \\
2008in	&NGC 4303   &1566   &--20.4     &0.83	&-15.40	&92.20	&0.23	&6903\\
\hline                      
\hline                      
\end{tabular}             
\label{snlist}
\end{table*}

\begin{table*}
\centering
\setcounter{table}{0}
\begin{tabular}[t]{cccccccccc}
\hline
SN & Host galaxy & $V_{\rm r}$(km s$^{-1}$) & Host $M_{\rm B}$ & $s_{\rm 2}$(mag 100d$^{-1}$) & $M_{\rm max}$(mag) & OPTd(d) & a/e & \ha$_{\rm vel}$(km s$^{-1}$) \\
\hline	
\hline
2009A	&anon&5160&$\cdots$&$\cdots$		&$\cdots$		&$\cdots$       &$\cdots$       &$\cdots$\\
2009N	&NGC 4487   &1034       &--20.2     &0.34	&--15.25	&89.50	&0.41	&5348\\
2009aj	&ESO 221-G18&2844&--19.1&$\cdots$		&$\cdots$		&$\cdots$       &$\cdots$       &$\cdots$\\
2009ao	&NGC 2939   &3339   &--20.5     &--0.01	&-15.79	&41.71	&	$\cdots$    &$\cdots$\\
2009au	&ESO 443-21 &2819   &--19.9     &3.04	&-16.34	&$\cdots$	&$\cdots$	    &$\cdots$\\
2009bu	&NGC 7408   &3494   &--20.9     &0.18	&-16.05	&$\cdots$	&0.5	&5934\\
2009bz	&UGC 9814   &3231   &--19.1     &0.50	&-16.46	&$\cdots$	&$\cdots$	    &  $\cdots$  \\
\hline                      
\hline                      
\end{tabular}       
\caption{SN names followed by their respective
host galaxies are listed in columns one and two. These are followed by the host galaxy recession velocity
(taken from NED: http://ned.ipac.caltech.edu/) in column 3, and host galaxy absolute $B$-band magnitude 
(taken from HyperLeda: http://leda.univ-lyon1.fr/) in column 4. We then list SN $V$-band photometric and
spectroscopic \ha\ parameters: $s_2$ the `plateau' decline rate, M$_{max}$
the absolute magnitude at maximum light, OPTd the optically thick phase duration,  
a/e the ratio of pEWs of \ha\ absorption to emission, and \ha$_{\rm vel}$ the FWHM velocity of \ha\ (with the latter two
measured at a common epoch), 
in columns 5, 6, 7, 8, and 9 respectively. The reader is referred to A14 and \cite{gut14} for more details of those
measurements. (It is important to note that the SN magnitudes we use within this analysis have not been corrected for 
host galaxy extinction.) * labels SNe~IIn, while ** labels SNe~IIb.}          
\end{table*}

\section{\hii-region abundances and SN pEWs}
SN~II host \hii-region abundances and measured SN pEWs are listed in
Table ~\ref{hiilist}.

\clearpage

\begin{table*}
\centering
\caption{\hii-region abundances and SN pEWs}
\begin{tabular}[t]{ccccccc}
\hline
SN & \hii\ distance (kpc) & M13 N2 (dex) & M13 O3N2 (dex) & PP04 N2 (dex) & PP04 O3N2 (dex) &pEW at 50d (\AA)\\
\hline	
\hline
1986L*	&0.74	&8.46$^{+0.02}_{-0.02}$&8.39$^{+0.02}_{-0.02}$&8.52$^{+0.04}_{-0.04}$&8.52$^{+0.03}_{-0.03}$&10.85$\pm$1.67\\
1990E*	&5.08	&8.48$^{+0.02}_{-0.02}$&$\cdots$&8.57$^{+0.04}_{-0.04}$&$\cdots$&$\cdots$	\\		
1990K*	&3.11	&8.51$^{+0.02}_{-0.02}$&8.47$^{+0.02}_{-0.02}$&8.63$^{+0.04}_{-0.05}$&8.63$^{+0.03}_{-0.03}$&10.48$\pm$0.48\\
1991al*	&3.39	&8.43$^{+0.02}_{-0.02}$&8.57$^{+0.02}_{-0.02}$&8.47$^{+0.03}_{-0.03}$&8.79$^{+0.03}_{-0.03}$&13.57$\pm$0.83\\
1992af*	&0.08	&8.36$^{+0.02}_{-0.02}$&8.29$^{+0.02}_{-0.02}$&8.37$^{+0.03}_{-0.03}$&8.36$^{+0.03}_{-0.03}$&$\cdots$	\\
1992am*	&0.49	&8.53$^{+0.02}_{-0.02}$&8.51$^{+0.02}_{-0.02}$&8.66$^{+0.05}_{-0.05}$&8.69$^{+0.03}_{-0.03}$&$\cdots$	\\	
1992ad	&2.66	&8.52$^{+0.05}_{-0.05}$&$\cdots$&8.65$^{+0.11}_{-0.12}$&$\cdots$&$\cdots$	\\
1992ba*	&0.27	&8.53$^{+0.02}_{-0.02}$&$\cdots$&8.67$^{+0.04}_{-0.05}$&$\cdots$&20.19$\pm$0.78\\
1993A	&0.03	&8.50$^{+0.04}_{-0.04}$&$\cdots$&8.60$^{+0.08}_{-0.08}$&$\cdots$&$\cdots$	\\
1993K	&0.32	&8.56$^{+0.03}_{-0.03}$&8.56$^{+0.03}_{-0.04}$&8.75$^{+0.07}_{-0.07}$&8.76$^{+0.05}_{-0.05}$&11.43$\pm$0.50\\	
1993S	&0.00	&8.60$^{+0.02}_{-0.02}$&$\cdots$&8.87$^{+0.06}_{-0.06}$&$\cdots$&17.01$\pm$0.06\\
1999br*	&0.35	&8.47$^{+0.03}_{-0.02}$&$\cdots$&8.55$^{+0.05}_{-0.05}$&$\cdots$&25.89$\pm$1.06\\
1999ca*	&1.84	&8.41$^{+0.02}_{-0.02}$&8.35$^{+0.02}_{-0.02}$&8.45$^{+0.03}_{-0.03}$&8.45$^{+0.03}_{-0.03}$&17.57$\pm$0.25\\
1999cr*	&0.64	&8.08$^{+0.02}_{-0.02}$&8.09$^{+0.02}_{-0.02}$&8.11$^{+0.02}_{-0.02}$&8.06$^{+0.03}_{-0.03}$&13.72$\pm$2.81\\	
1999eg	&0.93	&8.66$^{+0.05}_{-0.05}$&$\cdots$&9.03$^{+0.15}_{-0.17}$&$\cdots$&$\cdots$	\\
1999em*	&0.23	&8.51$^{+0.03}_{-0.03}$&$\cdots$&8.63$^{+0.06}_{-0.06}$&$\cdots$&22.59$\pm$1.07\\
2002ew	&1.09	&8.44$^{+0.05}_{-0.04}$&8.32$^{+0.05}_{-0.04}$&8.49$^{+0.08}_{-0.08}$&8.41$^{+0.07}_{-0.06}$&8.88$\pm$4.70\\
2002fa	&4.18	&8.55$^{+0.06}_{-0.05}$&$\cdots$&8.71$^{+0.13}_{-0.13}$&$\cdots$&15.11$\pm$2.34\\	
2002gd	&0.00	&8.49$^{+0.05}_{-0.05}$&8.36$^{+0.03}_{-0.03}$&8.58$^{+0.09}_{-0.10}$&8.47$^{+0.05}_{-0.05}$&22.51$\pm$3.26\\
2002gw*	&2.07	&8.24$^{+0.02}_{-0.02}$&8.22$^{+0.02}_{-0.02}$&8.24$^{+0.02}_{-0.02}$&8.27$^{+0.03}_{-0.03}$&19.91$\pm$1.04\\
2002hj	&2.65	&8.41$^{+0.01}_{-0.01}$&$\cdots$&8.45$^{+0.02}_{-0.02}$&$\cdots$&17.13$\pm$2.68\\
2002hx	&8.91	&8.55$^{+0.05}_{-0.05}$&8.49$^{+0.06}_{-0.07}$&8.72$^{+0.12}_{-0.12}$&8.66$^{+0.09}_{-0.10}$&21.04$\pm$1.06\\	
2002ig	&0.00	&8.34$^{+0.11}_{-0.09}$&$\cdots$&8.34$^{+0.11}_{-0.12}$&$\cdots$&$\cdots$	\\
2003B* 	&3.80	&8.55$^{+0.02}_{-0.02}$&8.57$^{+0.02}_{-0.02}$&8.72$^{+0.05}_{-0.05}$&8.79$^{+0.03}_{-0.03}$&24.55$\pm$0.91\\
2003E*	&0.28	&8.12$^{+0.02}_{-0.02}$&8.20$^{+0.02}_{-0.02}$&8.14$^{+0.02}_{-0.02}$&8.23$^{+0.03}_{-0.03}$&11.15$\pm$3.90\\
2003T*	&2.29	&8.51$^{+0.02}_{-0.02}$&8.49$^{+0.02}_{-0.02}$&8.64$^{+0.05}_{-0.05}$&8.66$^{+0.03}_{-0.03}$&23.97$\pm$0.58\\
2003bl	&2.26	&8.54$^{+0.01}_{-0.01}$&8.59$^{+0.01}_{-0.01}$&8.70$^{+0.02}_{-0.02}$&8.81$^{+0.02}_{-0.02}$&23.93$\pm$1.87\\
2003bn	&0.00	&8.52$^{+0.03}_{-0.03}$&$\cdots$&8.64$^{+0.06}_{-0.06}$&$\cdots$&15.08$\pm$3.31\\	
2003ci*	&5.18	&8.50$^{+0.02}_{-0.02}$&8.56$^{+0.02}_{-0.02}$&8.61$^{+0.04}_{-0.04}$&8.76$^{+0.03}_{-0.03}$&15.92$\pm$0.94\\
2003cn	&2.61	&8.38$^{+0.02}_{-0.02}$&8.35$^{+0.02}_{-0.02}$&8.40$^{+0.03}_{-0.03}$&8.45$^{+0.03}_{-0.03}$&16.93$\pm$2.54\\
2003cx	&0.48	&8.03$^{+0.06}_{-0.05}$&8.06$^{+0.03}_{-0.03}$&8.06$^{+0.06}_{-0.04}$&8.03$^{+0.05}_{-0.04}$&11.49$\pm$1.58\\
2003dq	&0.50	&8.15$^{+0.03}_{-0.03}$&8.15$^{+0.02}_{-0.02}$&8.17$^{+0.03}_{-0.03}$&8.15$^{+0.03}_{-0.03}$&$\cdots$	\\	
2003ef*	&1.62	&8.53$^{+0.02}_{-0.02}$&8.59$^{+0.02}_{-0.02}$&8.66$^{+0.04}_{-0.05}$&8.81$^{+0.03}_{-0.03}$&22.29$\pm$0.61\\
2003eg	&10.01	&8.57$^{+0.05}_{-0.04}$&$\cdots$&8.77$^{+0.11}_{-0.13}$&$\cdots$&$\cdots$	\\
2003ej	&0.13	&8.34$^{+0.04}_{-0.03}$&8.27$^{+0.02}_{-0.02}$&8.34$^{+0.04}_{-0.04}$&8.34$^{+0.03}_{-0.03}$&10.43$\pm$0.50\\	
2003fb*	&4.52	&8.49$^{+0.02}_{-0.02}$&8.46$^{+0.03}_{-0.03}$&8.58$^{+0.05}_{-0.05}$&8.62$^{+0.04}_{-0.04}$&19.23$\pm$1.51\\
2003gd*	&0.20	&8.45$^{+0.04}_{-0.04}$&8.45$^{+0.05}_{-0.06}$&8.50$^{+0.06}_{-0.07}$&8.61$^{+0.08}_{-0.09}$&30.00$\pm$0.50\\
2003hd*	&1.31	&8.46$^{+0.02}_{-0.02}$&8.38$^{+0.02}_{-0.02}$&8.52$^{+0.04}_{-0.04}$&8.50$^{+0.03}_{-0.03}$&15.34$\pm$1.37\\
2003hk*	&13.43	&8.51$^{+0.02}_{-0.02}$&8.44$^{+0.02}_{-0.02}$&8.62$^{+0.04}_{-0.05}$&8.60$^{+0.03}_{-0.03}$&18.43$\pm$1.16\\	
2003hl*	&1.00	&8.53$^{+0.02}_{-0.02}$&8.63$^{+0.02}_{-0.02}$&8.66$^{+0.04}_{-0.05}$&8.87$^{+0.03}_{-0.03}$&21.85$\pm$1.96\\
2003hn*	&3.30	&8.42$^{+0.02}_{-0.02}$&8.35$^{+0.02}_{-0.02}$&8.46$^{+0.03}_{-0.03}$&8.46$^{+0.03}_{-0.03}$&17.43$\pm$0.27\\
2003ho	&1.22	&8.54$^{+0.06}_{-0.05}$&$\cdots$&8.69$^{+0.13}_{-0.13}$&$\cdots$&16.94$\pm$0.65\\
2003ib	&1.64	&8.56$^{+0.03}_{-0.02}$&$\cdots$&8.74$^{+0.06}_{-0.06}$&$\cdots$&14.77$\pm$0.01\\	
2003ip	&3.46	&8.54$^{+0.06}_{-0.05}$&$\cdots$&8.71$^{+0.13}_{-0.14}$&$\cdots$&9.21$\pm$2.26\\
2003iq	&0.00	&8.51$^{+0.02}_{-0.02}$&$\cdots$&8.64$^{+0.04}_{-0.04}$&$\cdots$&19.81$\pm$0.67\\
2004dy	&1.31	&8.62$^{+0.01}_{-0.01}$&$\cdots$&8.91$^{+0.03}_{-0.04}$&$\cdots$&$\cdots$	\\	
2004ej	&0.78	&8.55$^{+0.04}_{-0.04}$&8.57$^{+0.04}_{-0.04}$&8.71$^{+0.10}_{-0.11}$&8.79$^{+0.06}_{-0.06}$&20.74$\pm$0.46\\
2004er	&0.52	&8.47$^{+0.01}_{-0.01}$&8.41$^{+0.01}_{-0.01}$&8.54$^{+0.02}_{-0.02}$&8.55$^{+0.02}_{-0.02}$&13.58$\pm$1.95\\
2004fb	&0.00	&8.56$^{+0.01}_{-0.01}$&$\cdots$&8.74$^{+0.02}_{-0.03}$&$\cdots$&15.48$\pm$0.50\\
2004fc	&0.00	&8.54$^{+0.02}_{-0.02}$&8.51$^{+0.03}_{-0.03}$&8.71$^{+0.05}_{-0.06}$&8.70$^{+0.05}_{-0.05}$&16.62$\pm$4.04\\	
2004fx	&5.05	&8.43$^{+0.05}_{-0.04}$&$\cdots$&8.47$^{+0.08}_{-0.08}$&$\cdots$&18.75$\pm$1.39\\
2005J	&1.70	&8.59$^{+0.03}_{-0.03}$&$\cdots$&8.81$^{+0.09}_{-0.09}$&$\cdots$&15.99$\pm$2.08\\
2005K	&0.14	&8.42$^{+0.05}_{-0.04}$&$\cdots$&8.46$^{+0.07}_{-0.07}$&$\cdots$&20.15$\pm$0.50\\
\hline                                                  
\hline                          
\end{tabular}                    
\end{table*}

\begin{table*}                   
\centering
\setcounter{table}{1}
\begin{tabular}[t]{ccccccc}
\hline
SN & \hii\ distance (kpc) & M13 N2 (dex) & M13 O3N2 (dex) & PP04 N2 (dex) & PP04 O3N2 (dex) &pEW at 50d (\AA)\\
\hline	
\hline
2005Z	&-0.01	&8.55$^{+0.02}_{-0.02}$&$\cdots$&8.71$^{+0.06}_{-0.06}$&$\cdots$&12.36	0.75\\
2005af	&0.55	&8.53$^{+0.07}_{-0.06}$&$\cdots$&8.66$^{+0.14}_{-0.15}$&$\cdots$&$\cdots$	\\
2005an	&0.00	&8.53$^{+0.03}_{-0.02}$&$\cdots$&8.67$^{+0.06}_{-0.06}$&$\cdots$&15.07$\pm$1.26\\
2005dk	&1.82	&8.54$^{+0.03}_{-0.03}$&$\cdots$&8.69$^{+0.07}_{-0.07}$&$\cdots$&13.71$\pm$0.37\\	
2005dn	&0.56	&8.15$^{+0.03}_{-0.03}$&$\cdots$&8.17$^{+0.02}_{-0.03}$&$\cdots$&4.05$\pm$0.65\\
2005dt	&6.84	&8.44$^{+0.08}_{-0.06}$&$\cdots$&8.49$^{+0.12}_{-0.12}$&$\cdots$&$\cdots$	\\
2005dw	&0.00	&8.53$^{+0.01}_{-0.01}$&8.51$^{+0.01}_{-0.01}$&8.68$^{+0.01}_{-0.01}$&8.69$^{+0.02}_{-0.02}$&$\cdots$	\\
2005dx	&0.00	&8.52$^{+0.03}_{-0.03}$&$\cdots$&8.65$^{+0.07}_{-0.07}$&$\cdots$&$\cdots$	\\	
2005dz	&2.21	&8.54$^{+0.04}_{-0.04}$&$\cdots$&8.69$^{+0.08}_{-0.10}$&$\cdots$&35.56$\pm$3.05\\
2005es	&0.00	&8.60$^{+0.06}_{-0.06}$&$\cdots$&8.85$^{+0.16}_{-0.18}$&$\cdots$&$\cdots$	\\
2005gk	&0.00	&8.51$^{+0.04}_{-0.04}$&$\cdots$&8.62$^{+0.07}_{-0.08}$&$\cdots$&$\cdots$	\\
2005gz	&0.00	&8.55$^{+0.01}_{-0.01}$&8.55$^{+0.03}_{-0.03}$&8.71$^{+0.03}_{-0.03}$&8.75$^{+0.04}_{-0.05}$&$\cdots$	\\
2005kh	&3.36	&8.55$^{+0.05}_{-0.05}$&$\cdots$&8.71$^{+0.11}_{-0.15}$&$\cdots$&$\cdots$	\\
2005me	&3.17	&8.48$^{+0.03}_{-0.03}$&$\cdots$&8.56$^{+0.06}_{-0.06}$&$\cdots$&$\cdots$	\\
2006Y	&0.00	&8.28$^{+0.04}_{-0.04}$&8.21$^{+0.03}_{-0.03}$&8.28$^{+0.04}_{-0.04}$&8.25$^{+0.05}_{-0.04}$&6.74$\pm$1.18\\
2006ai	&0.00	&8.51$^{+0.03}_{-0.03}$&8.49$^{+0.02}_{-0.02}$&8.62$^{+0.06}_{-0.06}$&8.66$^{+0.03}_{-0.03}$&15.43$\pm$1.26\\
2006bc	&0.00	&8.57$^{+0.03}_{-0.03}$&8.53$^{+0.04}_{-0.04}$&8.77$^{+0.08}_{-0.09}$&8.72$^{+0.06}_{-0.06}$&$\cdots$	\\
2006be	&0.04	&8.57$^{+0.02}_{-0.02}$&8.35$^{+0.02}_{-0.02}$&8.76$^{+0.04}_{-0.04}$&8.46$^{+0.03}_{-0.03}$&15.72$\pm$0.71\\
2006bl	&0.00	&8.51$^{+0.03}_{-0.03}$&$\cdots$&8.62$^{+0.06}_{-0.06}$&$\cdots$&$\cdots$	\\
2006it	&0.00	&8.61$^{+0.02}_{-0.02}$&$\cdots$&8.89$^{+0.05}_{-0.05}$&$\cdots$&$\cdots$	\\
2006iw	&5.10	&8.55$^{+0.01}_{-0.01}$&$\cdots$&8.72$^{+0.04}_{-0.04}$&$\cdots$&13.25$\pm$0.50\\
2006ms	&0.00	&8.57$^{+0.03}_{-0.03}$&$\cdots$&8.78$^{+0.07}_{-0.07}$&$\cdots$&$\cdots$	\\
2006qq	&0.00	&8.52$^{+0.03}_{-0.02}$&8.54$^{+0.04}_{-0.04}$&8.65$^{+0.05}_{-0.05}$&8.74$^{+0.06}_{-0.06}$&$\cdots$	\\
2006qr	&1.31	&8.63$^{+0.04}_{-0.04}$&$\cdots$&8.95$^{+0.12}_{-0.13}$&$\cdots$&24.84$\pm$2.26\\
2007W	&0.00	&8.47$^{+0.06}_{-0.05}$&8.45$^{+0.06}_{-0.06}$&8.54$^{+0.10}_{-0.10}$&8.60$^{+0.09}_{-0.09}$&24.22$\pm$1.23\\
2007X	&4.16	&8.60$^{+0.05}_{-0.05}$&$\cdots$&8.85$^{+0.13}_{-0.15}$&$\cdots$&13.81$\pm$0.99\\
2007Z	&0.00	&8.56$^{+0.02}_{-0.02}$&$\cdots$&8.75$^{+0.06}_{-0.06}$&$\cdots$&$\cdots$	\\
2007am	&0.00	&8.53$^{+0.01}_{-0.01}$&8.58$^{+0.01}_{-0.01}$&8.68$^{+0.01}_{-0.01}$&8.81$^{+0.02}_{-0.02}$&$\cdots$	\\
2007aa	&0.00	&8.76$^{+0.01}_{-0.01}$&$\cdots$&9.45$^{+0.04}_{-0.04}$&$\cdots$&23.55$\pm$0.67\\
2007ab	&9.88	&8.58$^{+0.06}_{-0.06}$&$\cdots$&8.80$^{+0.16}_{-0.17}$&$\cdots$&17.35$\pm$1.54\\
2007av	&0.00	&8.58$^{+0.02}_{-0.02}$&8.51$^{+0.03}_{-0.03}$&8.80$^{+0.06}_{-0.07}$&8.69$^{+0.04}_{-0.04}$&22.50$\pm$0.14\\
2007ay	&0.26	&8.58$^{+0.02}_{-0.02}$&$\cdots$&8.80$^{+0.05}_{-0.05}$&$\cdots$&$\cdots$	\\
2007hm	&10.99	&8.47$^{+0.02}_{-0.02}$&8.44$^{+0.05}_{-0.05}$&8.55$^{+0.03}_{-0.03}$&8.59$^{+0.08}_{-0.08}$&12.85$\pm$1.00\\
2007il	&0.00	&8.49$^{+0.02}_{-0.02}$&8.40$^{+0.02}_{-0.02}$&8.57$^{+0.04}_{-0.04}$&8.54$^{+0.03}_{-0.02}$&14.23$\pm$2.22\\
2007it	&1.47	&8.58$^{+0.02}_{-0.02}$&$\cdots$&8.80$^{+0.07}_{-0.07}$&$\cdots$&$\cdots$	\\
2007oc	&1.54	&8.36$^{+0.02}_{-0.02}$&8.23$^{+0.02}_{-0.02}$&8.37$^{+0.02}_{-0.02}$&8.28$^{+0.03}_{-0.03}$&10.01$\pm$1.77\\
2007sq	&3.08	&8.52$^{+0.03}_{-0.03}$&$\cdots$&8.64$^{+0.07}_{-0.07}$&$\cdots$&8.75$\pm$0.50\\
2008F	&0.05	&8.59$^{+0.06}_{-0.05}$&$\cdots$&8.81$^{+0.15}_{-0.15}$&$\cdots$&$\cdots$	\\
2008H	&0.13	&8.63$^{+0.01}_{-0.01}$&$\cdots$&8.93$^{+0.04}_{-0.04}$&$\cdots$&22.00$\pm$2.40\\
2008M	&0.00	&8.43$^{+0.03}_{-0.03}$&8.34$^{+0.03}_{-0.03}$&8.48$^{+0.05}_{-0.05}$&8.44$^{+0.05}_{-0.05}$&20.36$\pm$3.95\\
2008N	&0.00	&8.58$^{+0.02}_{-0.02}$&$\cdots$&8.81$^{+0.06}_{-0.06}$&$\cdots$&$\cdots$	\\
2008W	&0.00	&8.60$^{+0.04}_{-0.04}$&8.48$^{+0.04}_{-0.05}$&8.86$^{+0.11}_{-0.13}$&8.66$^{+0.07}_{-0.07}$&17.80$\pm$2.61\\2008ag	&4.24	&8.50$^{+0.01}_{-0.01}$&$\cdots$&8.61$^{+0.02}_{-0.02}$&$\cdots$&23.88$\pm$0.73\\
2008aw	&0.49	&8.62$^{+0.01}_{-0.01}$&8.51$^{+0.02}_{-0.02}$&8.91$^{+0.04}_{-0.04}$&8.70$^{+0.03}_{-0.03}$&12.47$\pm$0.60\\
2008bh	&0.51	&8.58$^{+0.04}_{-0.04}$&$\cdots$&8.80$^{+0.09}_{-0.10}$&$\cdots$&16.15$\pm$0.50\\
2008bk	&0.17	&8.49$^{+0.06}_{-0.05}$&$\cdots$&8.58$^{+0.11}_{-0.12}$&$\cdots$&24.12$\pm$0.83\\
2008bm	&0.00	&8.33$^{+0.02}_{-0.02}$&8.24$^{+0.02}_{-0.02}$&8.33$^{+0.02}_{-0.02}$&8.28$^{+0.03}_{-0.02}$&10.32$\pm$0.81\\
2008bp	&2.39	&8.54$^{+0.06}_{-0.05}$&$\cdots$&8.68$^{+0.14}_{-0.14}$&$\cdots$&34.47$\pm$6.53\\
2008br	&1.17	&8.46$^{+0.02}_{-0.02}$&8.40$^{+0.02}_{-0.02}$&8.53$^{+0.04}_{-0.04}$&8.54$^{+0.03}_{-0.03}$&23.53$\pm$4.98\\
2008bu	&0.00	&8.26$^{+0.02}_{-0.02}$&$\cdots$&8.26$^{+0.02}_{-0.02}$&$\cdots$&$\cdots$	\\
2008fq	&0.00	&8.57$^{+0.04}_{-0.04}$&$\cdots$&8.78$^{+0.09}_{-0.10}$&$\cdots$&$\cdots$	\\
2008ga	&16.18	&8.50$^{+0.07}_{-0.06}$&$\cdots$&8.61$^{+0.13}_{-0.14}$&$\cdots$&17.00	2.30\\
2008gi	&0.00	&8.53$^{+0.02}_{-0.02}$&8.45$^{+0.03}_{-0.04}$&8.68$^{+0.04}_{-0.05}$&8.61$^{+0.05}_{-0.06}$&16.38$\pm$1.75\\
2008gq	&7.46	&8.44$^{+0.02}_{-0.02}$&8.41$^{+0.03}_{-0.03}$&8.48$^{+0.04}_{-0.04}$&8.55$^{+0.04}_{-0.04}$&$\cdots$	\\
2008gr	&1.97	&8.46$^{+0.03}_{-0.03}$&$\cdots$&8.52$^{+0.05}_{-0.06}$&$\cdots$&6.26$\pm$1.11\\
2008ho	&0.00	&8.30$^{+0.01}_{-0.01}$&8.26$^{+0.01}_{-0.01}$&8.30$^{+0.01}_{-0.01}$&8.33$^{+0.02}_{-0.02}$&$\cdots$	\\
\hline                      
\hline                   
\end{tabular}               
\end{table*}

\begin{table*}
\centering
\setcounter{table}{1}
\begin{tabular}[t]{ccccccc}
\hline
SN & \hii\ distance (kpc) & M13 N2 (dex) & M13 O3N2 (dex) & PP04 N2 (dex) & PP04 O3N2 (dex) &pEW at 50d (\AA)\\
\hline	
\hline
2008if	&0.37	&8.62$^{+0.17}_{-0.11}$&$\cdots$&8.90$^{+0.40}_{-0.38}$&$\cdots$&9.67$\pm$1.17\\
2008il	&1.88	&8.37$^{+0.03}_{-0.03}$&$\cdots$&8.38$^{+0.04}_{-0.04}$&$\cdots$&$\cdots$	\\
2008in	&0.34	&8.61$^{+0.01}_{-0.01}$&8.53$^{+0.01}_{-0.01}$&8.88$^{+0.02}_{-0.02}$&8.72$^{+0.02}_{-0.02}$&24.62$\pm$0.77\\
2009A	&0.00	&8.22$^{+0.03}_{-0.03}$&8.19$^{+0.02}_{-0.02}$&8.22$^{+0.02}_{-0.02}$&8.22$^{+0.03}_{-0.02}$&$\cdots$	\\
2009N	&0.23	&8.48$^{+0.02}_{-0.02}$&$\cdots$&8.57$^{+0.04}_{-0.05}$&$\cdots$&26.60$\pm$0.67\\
2009aj	&3.49	&8.29$^{+0.03}_{-0.03}$&8.31$^{+0.04}_{-0.04}$&8.29$^{+0.03}_{-0.03}$&8.40$^{+0.07}_{-0.06}$&9.02$\pm$0.90\\
2009ao	&0.00	&8.55$^{+0.02}_{-0.02}$&$\cdots$&8.71$^{+0.05}_{-0.05}$&$\cdots$&18.87$\pm$0.39\\
2009au	&0.00	&8.56$^{+0.03}_{-0.03}$&$\cdots$&8.76$^{+0.08}_{-0.08}$&$\cdots$&14.65$\pm$2.63\\
2009bu	&0.42	&8.42$^{+0.02}_{-0.02}$&$\cdots$&8.45$^{+0.03}_{-0.04}$&$\cdots$&14.19$\pm$0.31\\
2009bz	&3.44	&8.43$^{+0.03}_{-0.03}$&$\cdots$&8.48$^{+0.05}_{-0.05}$&$\cdots$&$\cdots$	\\
\hline                      
\hline                      
\end{tabular}         
\caption{In the first column we list the SN name. In column 2 the distance of the 
spectral extraction region from the explosion site is given. In columns 3 and 4 we list the oxygen abundances as calculated
by the M13 N2 and O3N2 diagnostics respectively. These are followed by the the oxygen abundances as calculated
by the \cite{pet04} N2 and O3N2 diagnostics in columns 5 and 6. Finally in column 7 we list the \feii\ 5018\,\AA\ pEWs as estimated
at
50\,d. Errors on pEWs are those obtained from the fitting process to the 
sample of pEWs for each SN. In the case of a straight-line fit to two data points we set the minimum error
to 0.5\AA. The inferred abundances of SN~2005dn are upper limits as \nii\ was not detected in the spectra. To estimate 
abundances a 3$\sigma$ \nii\ flux upper limit was calculated using an estimation of the RMS noise in the spectrum close in wavelength
to the
spectral line. Removing this SN from our correlations has a negligible effect on our results and conclusions.
The SNe where abundance measurements are taken from \cite{and10} are indicated by * next to the SN name.}   
\label{hiilist}   
\end{table*}

\end{document}